\pdfoutput=1
\documentclass[pdflatex,sn-nature]{sn-jnl}
\usepackage{graphicx}%
\usepackage[font=small, justification=justified]{caption}
\usepackage{multirow}%
\usepackage{amsmath,amssymb,amsfonts}%
\usepackage{amsthm}%
\usepackage{mathrsfs}%
\usepackage{color}
\usepackage{xcolor}%
\usepackage{textcomp}%
\usepackage{manyfoot}%
\usepackage{booktabs}%
\usepackage{algpseudocode}%
\usepackage{listings}%
\usepackage{epstopdf}
\hypersetup{
    colorlinks=true,
    linkcolor=cyan,
    filecolor=green,      
    urlcolor=blue,
    }
\usepackage{geometry}
\usepackage{lipsum}
\usepackage{relsize}
\usepackage{array}
\usepackage{float}

\usepackage[final]{pdfpages}

\begin{document}

\title[Article Title]{Machine Learning, Density Functional Theory, and Experiments to Understand the Photocatalytic Reduction of CO$_2$ by CuPt/TiO$_2$}

\author[1,2]{\fnm{Vaidish} \sur{Sumaria}}
\author*[1]{\fnm{Takat B.} \sur{Rawal}}\email{takat.rawal@genmat.xyz}
\author[1]{\fnm{Young Feng} \sur{Li}}
\author[1]{\fnm{David} \sur{Sommer}}
\author[1]{\fnm{Jake} \sur{Vikoren}}
\author[1]{\fnm{Robert J.} \sur{Bondi}}
\author[1,3]{\fnm{Matthias} \sur{Rupp}}
\author[1]{\fnm{Amrit} \sur{Prasad}}
\author[1]{\fnm{Deeptanshu} \sur{Prasad}}
\affil*[1]{\orgname{Quantum Generative Materials (GenMat)}, \orgaddress{\street{411 W. Monroe St}, \city{Austin}, \state{TX} \postcode{78704}, \country{USA}}}
\affil[2]{\orgname{Caminosoft Technologies Inc.}, \orgaddress{\street{1197 E Los Angeles Ave St C305}, \city{Simi Valley}, \state{CA} \postcode{93065}, \country{USA}}}
\affil[3]{\orgname{Luxembourg Institute of Science and Technology (LIST)}, \orgaddress{\state{Belvaux}, \country{Luxembourg}}}

\abstract{The photoconversion of CO$_2$ to hydrocarbons is a sustainable route to its transformation into value-added compounds and, thereby, crucial to mitigating the energy and climate crises. CuPt nanoparticles on TiO$_2$ surfaces have been reported to show promising photoconversion efficiency. For further progress, a mechanistic understanding of the catalytic properties of these CuPt/TiO$_2$ systems is vital. Here, we employ \textit{ab-initio} calculations, machine learning, and photocatalysis experiments to explore their configurational space and examine their reactivity and find that the interface plays a key role in stabilizing *CO$_2$, *CO, and other CH-containing intermediates, facilitating higher activity and selectivity for methane. A bias-corrected machine-learning interatomic potential trained on density functional theory data enables efficient exploration of the potential energy surfaces of numerous CO$_2$@CuPt/TiO$_2$ configurations via basin-hopping Monte Carlo simulations, greatly accelerating the study of these photocatalyst systems. Our simulations show that CO$_2$ preferentially adsorbs at the interface, with C atom bonded to a Pt site and one O atom occupying an O-vacancy site. The interface also promotes the formation of *CH and *CH$_2$ intermediates. For confirmation, we synthesize CuPt/TiO$_2$ samples with a variety of compositions and analyze their morphologies and compositions using scanning electron microscopy and energy-dispersive X-ray spectroscopy, and measure their photocatalytic activity. Our computational and experimental findings qualitatively agree and highlight the importance of interface design for selective conversion of CO$_2$ to hydrocarbons.}


\maketitle
\section{Introduction}\label{sec1}
CO$_2$ capture and its efficient conversion into value-added products are important steps to mitigating the energy and climate crises. CO$_2$-derived chemicals like polycarbonates, urea etc. and fuels like methane, alcohols, jet fuel etc. have industrial applications, and thus can provide routes for monetization. Owing to its natural abundance, low-operating costs, high-chemical stability, low toxicity, and environmental compatibility, TiO$_2$ and its derived materials have been extensively investigated over roughly seven decades, with several promising applications \cite{Haider_Jameel, Li_Li_Zuo, Mao_Sun, Wang_Qi} including catalysis \cite{Adachi_Brndiar, Bikondoa_Pang, Galhenage_Yan, Park_Ratliff, Yuan_Zhu} and photocatalysis.\cite{Gao_Wei, Li_Wang, Schneider_Matsuoka, Sorcar_Hwang_b} Using mechanistic insights to rationally optimize TiO$_2$-based materials is fundamental to enhance CO$_2$-capture efficiency and improve activity and selectivity for its conversion to hydrocarbons.

Despite the challenges of low-yield ($\sim \mu$ mol$^{-1}g^{-1}hr^{-1}$) for large-scale applications in photocatalytic CO$_2$ conversion to hydrocarbons, the activity and selectivity of TiO$_2$-based systems \cite{Lee_Jeong, Sorcar_Hwang_a, Sorcar_Hwang_b, Tasbihi_Fresno} are encouraging. Since pristine TiO$_2$ suﬀers from a low yield due to several reasons, including the maximum-visible light under-utilization,\cite{Etacheri_Valentin} large electron-hole separation,\cite{Gao_Wei} and wide band gaps $\sim$3.0 eV (rutile)\cite{Amtout_Leonelli} and 3.2 eV (anatase),\cite{Scanlon_Dunnill} several strategies have been tested insofar. Manipulating the geometric/electronic properties of TiO$_2$ via creation of oxygen vacancies or depositing metal (e.g. Cu/Pt) nanoparticles have potential to enhance the photocatalytic activity.\cite{Liu_Gao} 

On TiO$_2$ surfaces, CO$_2$ activation is a bottleneck step owing to a variety of reasons.\cite{Acharya_Camillone, Sorescu_Lee, Lin_Yoon, Zhao_Xu} For instance, the *CO$_2$ formation via absorption of photo-excited electrons: CO$_{2(g)}$ $\rightarrow$ *CO$_2$+e$^{–}$ is thermodynamically hindered whilst the *CHOO formation via absorption of two electrons: CO$_{2(g)}$+H $\rightarrow$ *CHOO+2e$^{–}$ is kinetically hindered.\cite{Zhao_Xu} The trivial charge transfer between *CO$_2$ and TiO$_2$ would result in weaker interaction.\cite{Acharya_Camillone, Sorescu_Lee, Lin_Yoon} Since CO$_2$ weakly interacts with defect-laden TiO$_2$ (see section S1 and Fig.S1 in supporting information) it cannot facilitate CO$_2$ activation without external driving force, thus necessitating an engineering of TiO$_2$ structures.

The manipulation of geometrical and electronic structures of TiO$_2$-based systems has remarkable impacts on their chemical activity. Under reducing conditions, a TiO$_2$ surface contains O vacancies which bolster metal-support interaction and enhance its reactivity.\cite{Liu_Zhao, Palfey_Rossman} The Cu-decorated TiO$_{2-x}$ promotes CO$_2$ conversion.\cite{Kattel_Yan_b} The suitable band realignment due to charge transfer,\cite{Kar_Zhang} enhances activity of Pt/TiO$_2$ for CO$_2$ conversion to CH$_4$.\cite{Kattel_Yan_a} The PtCu/TiO$_2$ shows better activity than Pt/TiO$_2$.\cite{Qi_Yang} CuPt/titania offers a good activity towards the photo-catalytic CO$_2$ conversion and higher selectivity for CH$_4$ (yield $\sim$ 92\%).\cite{Sorcar_Hwang_b} Due to a low percentage, the TiO$_2$(110) might be mostly exposed as a reactive facet. In this work, we model CuPt/TiO$_2$ systems by considering the rutile TiO$_2$(110) with O vacancy and 13-atom CuPt clusters.

For the accelerated discovery of materials for CO$_2$ reduction, various machine learning (ML) models have been utilized to study several systems \cite{Abraham_Pique, Mazheika_Wang, Mok_Li, Zhong_Tran} including (bare) semiconductor oxides.\cite{Mazheika_Wang} More recently, E(3)-equivariant neural network (ENN) is attracted attention owing to high data efficiency for ML models.\cite{Batzner_Musaelian, Musaelian_Batzner} The ML approach in combination with basin-hopping Monte Carlo (BHMC) method can provide an efficient way to tackle the combinatorial problem of composition of bimetallic Cu-Pt nanoclusters together with CO$_2$ adsorbates, relating to the large degrees of freedom due to several adsorption sites and the orientation of clusters and CO$_2$.

Whilst the \textit{ab-initio} methods offer atomistic insights into the potential energy surface (PES) of material systems but suffer from the high computational cost and inefficient scaling with system sizes, ML interatomic potentials (MLIPs) can provide substantial speed-ups at the expense of accuracy with linear scaling. ML can often reduce computational costs by orders of magnitude.\cite{Langer_Goeßmann_Rupp, Xie_Rupp} MLIPs have been improved drastically to achieve higher accuracy in energies ($<1$ meV/atom) and forces ($<$0.1 eV/\AA) and have been utilized in several applications, e.g. in understanding complex materials chemistry,\cite{Andrade_Ko, Gupta_Yang, Sumaria_Nguyen, Paleico_Behler, ko_Finkler, Sumaria_Sautet}, enumerating nanostructures configurations,\cite{Sumaria_Nguyen, Sun_Sautet} performing long-time MD simulations,\cite{Lim_Vandermause, Vandermause_Xie, Wen_Andrade} and exploring catalytic reactions.\cite{Jung_Sauerland, Bruix_Margraf, Schaaf_Fako, Miyazaki_Belthle} Nevertheless, MLIPs suffer from highly inhomogeneous feature-space sampling of training set (inherent biases). The inclusion of underrepresented configurations leads to significant errors since the data representing inhomogeneous local environments are overshadowed by core atoms.\cite{Jeong_Lee} To model CO$_2$@CuPt/TiO$_2$ systems, it thus requires an inherent-bias correction.

Herein, we perform the \textit{ab-initio} calculations of Cu$_{(13-n)}$Pt$_n$/TiO$_2$(110) model systems to generate the atom coordinates, energy, and forces for the initial training of MLIP based on ENN.\cite{Musaelian_Batzner} We modify the algorithm to reduce unwanted errors in under-represented configurations by using the Gaussian density function-based weighting scheme, improving MLIP reliability and transferability. We use the unbiased MLIP together with BHMC algorithm to enumerate several configurations of CO$_2$@Cu$_{(13-n)}$Pt$_n$/TiO$_2$ and find that CO$_2$ adsorbs at interfacial sites. CO$_2$ activation is then validated by DFT with calculations of adsorption energy, CO$_2$ bond length/angle changes, and charge transfer. Having examined the key reaction pathways for overall reaction between CO$_2$ and H$_2$O, we provide mechanistic insights into the role of CuPt/TiO$_2$ interface in CO$_2$ reduction activity. Our simulations results qualitatively agree with photocatalysis experiments.

\section{Results}\label{sec2}
\subsection{MLIP training and validation}\label{subsec2}
In Fig.1, we show the parity plot of MLIP-predicted against DFT-calculated energy and forces, along with the distribution of errors (inset). For the training set (8,646 structures), the root-mean-squared error (RMSE) and mean-absolute error (MAE) in energy are 0.63 and 0.46 meV/atom, respectively, while those in forces are 0.06 and 0.04 eV/\AA, respectively. The model is also tested using a validation set (9,633), resulting in the equivalent RMSE and MAE errors.

To obtain a better understanding of MLIP errors to learn \textit{ab-initio} data, we obtain decomposed-force parities for CO$_2$, CuPt cluster, and TiO$_2$ (Fig.S2). Even with less representation in the overall training data, force prediction errors for both CO$_2$ and nanocluster are comparable to the overall RMSE of force distribution (Fig.S2). The effect of weighted training has been discussed further in the supporting information (Fig.S3).

\begin{figure}[h]
\centering
\includegraphics[width=0.9\textwidth]{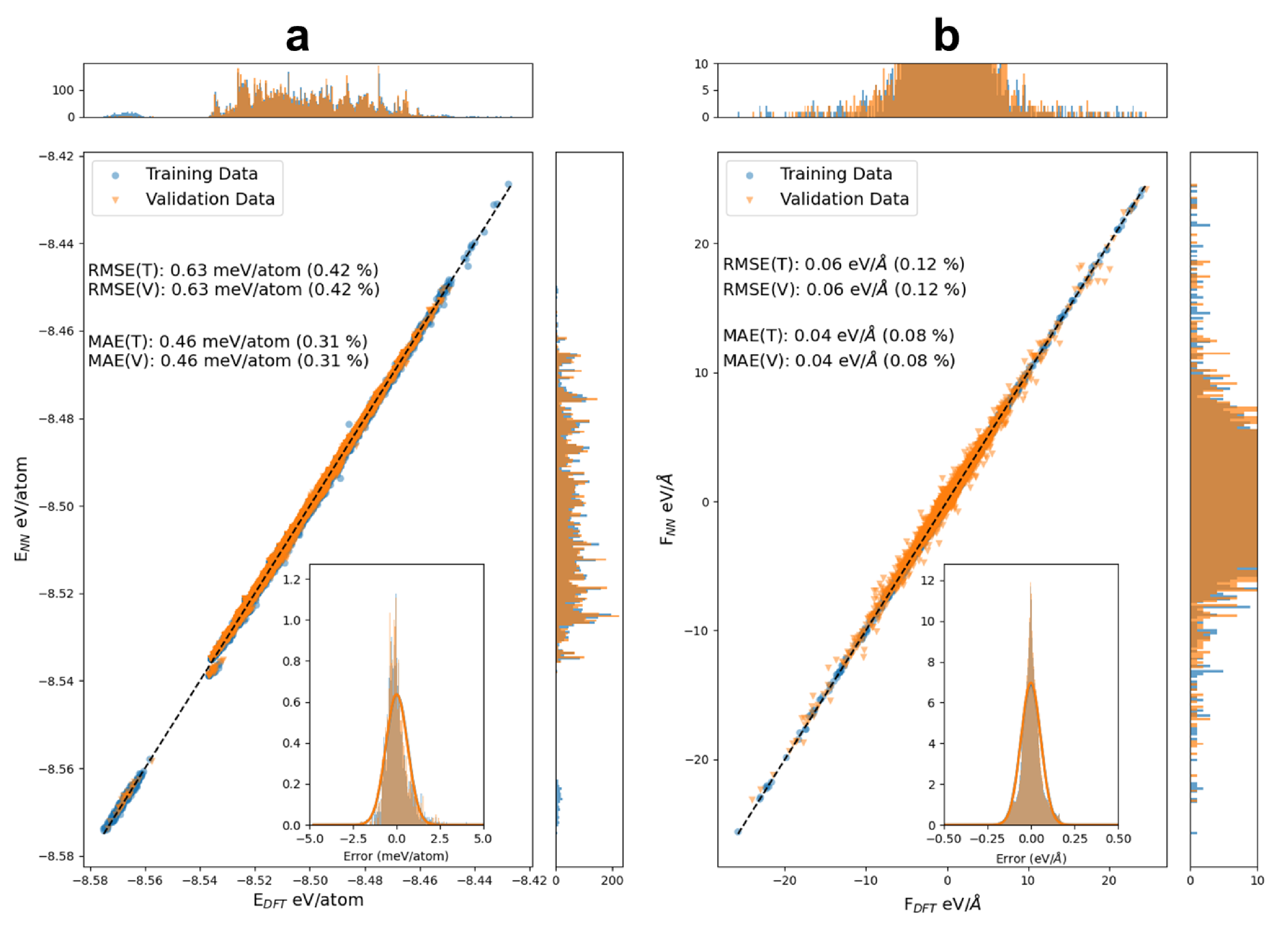}
\caption{Parity plots showing the performance of the MLIP for CO$_2$ adsorbed on Cu$_{(13-n)}$Pt$_n$/TiO$_2$ systems, $n=0, ...,13$, for a) energies and b) forces. The histograms on the margins show the corresponding distributions for training and validation data. Insets: Distribution of prediction errors; the solid lines indicate a fitted normal distribution.}\label{fig1}
\end{figure}

\begin{figure}[h]
\centering
\includegraphics[width=0.9\textwidth]{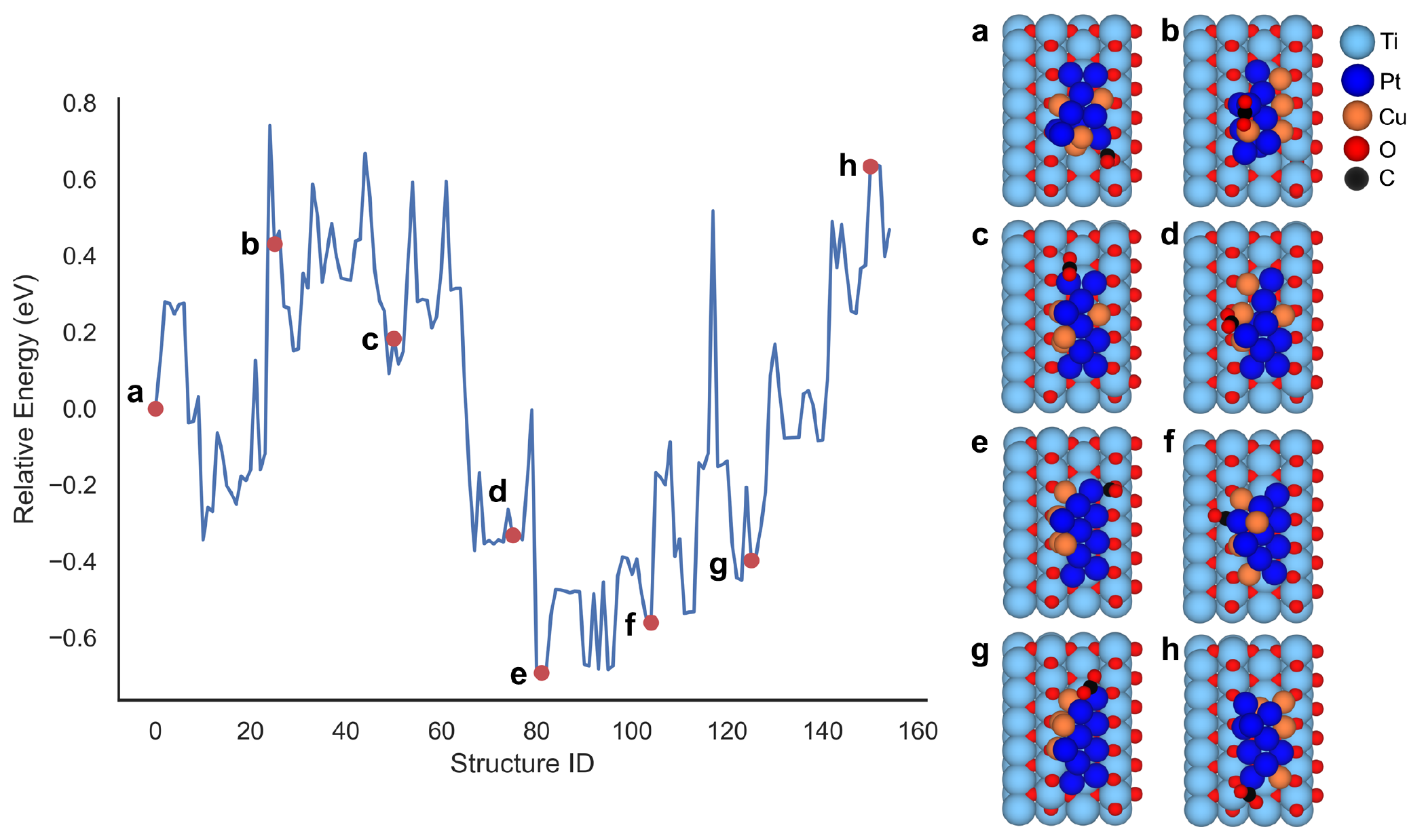}
\caption{Exploration of structures of CO$_2$@Cu$_{(13-n)}$Pt$_n$ systems via modified basin-hoping Monte Carlo, for example $n=9$. The relative energy (left) is given with respect to the energy of the initial seed configuration. Red dots represent minima of the potential energy surface. The corresponding structures are also shown (right, a-h). The configuration (e)  has the lowest energy.}\label{fig2}
\end{figure}

\subsection{Potential energy surface exploration for CO$_2$ adsorption on CuPt/TiO$_2$}\label{subsec2}
We perform BHMC+MLIP simulations to efficiently determine low-energy configurations of CO$_2$@Cu$_{(13-n)}$Pt$_n$/TiO$_2$. Analyzing the relative energy of CO$_2$@Cu$_{(13-n)}$Pt$_n$/TiO$_2$, e.g. $n=9$ (Fig.2), the BHMC-explored PES curve reveals that initially explored structures (b-c) are higher in energy which evolve towards minima region structures (e-f). If lower-energy structures are not found, the algorithm gradually accepts higher-energy configurations, moving to higher-energy basins (g-h).

The exploration yields the lowest energy basins with a common feature i.e. CO$_2$ adsorption occurs at CuPt/TiO$_2$ interface. We perform similar simulations with different Cu/Pt cluster compositions and discuss the minima CO$_2$ adsorption structures (section 2.4). We provide details of exploration procedure in SI (see section-S2.2). Despite MLIP's low validation errors, comparing structural stability within a small energy range is not viable. We adopt a 0.1 eV energy window, adding BHMC-generated structures with energy differences less than 0.1 eV to the low-energy ensemble. Subsequently, DFT relaxation is applied to this ensemble to eliminate the potential small MLIP errors.

The extremely large-configurational space is apparent from an illustration of PES exploration for a model system (Fig.2), whose configurations are found to be different cluster shapes with various arrangement of Pt/Cu atoms together with several possible adsorption sites for CO$_2$, including the CuPt/TiO$_2$ interfacial sites as well as different oxygen vacancy sites on TiO$_2$(110).

\subsection{Interaction of CuPt nanocluster with TiO$_2$ }\label{subsec2}
For CuPt composition, the cluster-support interaction is facilitated mostly by chemical bonds formed between Cu and bridging O atoms and bonds between Pt and bridging O and/or five-fold ($5f$) coordinated Ti atoms. The distributions of Cu-O bond lengths for the lowest-energy configurations (Fig.S4) and more details can be found in SI (see S2.3). Owing to the metal-support interaction, the Cu$_{(13-n)}$Pt$_n$ cluster adopts a hemi-spherical shape (Fig.3). Regardless of the composition, the supported CuPt cluster prefers a 3D geometrical shape.

\begin{figure}[h]
\centering
\includegraphics[width=0.9\textwidth]{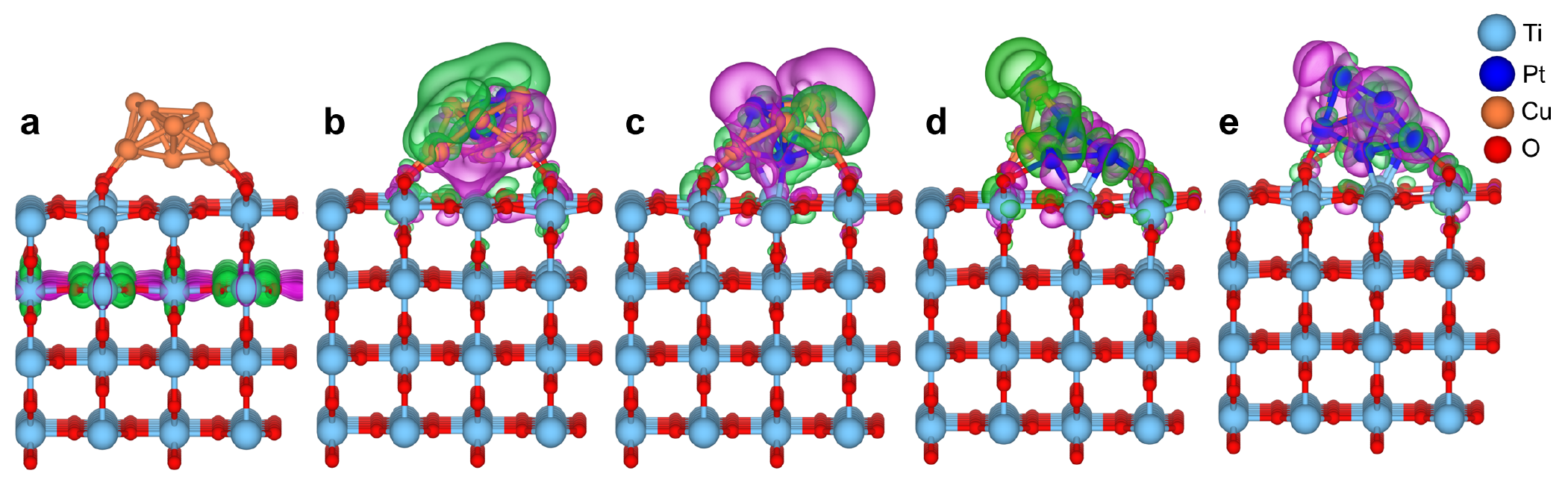}
\caption{Spatial localization of frontier electronic states corresponding to the topmost valence band calculated at the Brillouin zone center of the Cu$_{(13-n)}$Pt$_n$/TiO$_2$ systems for $n$ equal to: a) 0, b) 3, c) 5, d) 9, and e) 11. The green and pink iso-surfaces represent respectively the partial electron densities of +10$^{-4}$ $e/a_0^3$ and -10$^{-4}$ $e/a_0^3$, where $a_0$ is Bohr radius. The electronic states localize at the interface and around the Cu-Pt atoms.}\label{fig3}
\end{figure}

It is important to understand the distribution of CuPt/TiO$_2$ frontier states since the states spatially distributed around interfacial atoms would play an important role in ultrafast dynamics and thereby photo-induced reactions.\cite{Vaida_Rawal} For Cu$_{(13-n)}$Pt$_n$/TiO$_2$ ($n=0, 3, 5, 9, 11$), the frontier states are spattially localized at interface and around Cu-Pt atoms (Fig.3). Although the spatial localization is found to vary with Cu-Pt composition, the common feature is that there is some distribution of maximum valence band (VB) frontier states. In contrast, Cu$_{13}$/TiO$_2$ VB frontier states localize at the TiO$_2$ second layer (from the top). These results suggest that the CuPt/TiO$_2$ interface would be more active than Cu/TiO$_2$ interface.

\subsection{CO$_2$ interaction with CuPt/TiO$_2$}\label{subsec2}

We now examine the interaction of CO$_2$ with Cu$_{(13-n)}$Pt$_n$/TiO$_2$ via DFT by considering the lowest-minima configurations predicted by MLIP+BHMC simulations. For all CuPt compositions, we find that CO$_2$ adsorption is thermodynamically favorable, and CO$_2$ is activated (Table 1). Regardless of the CuPt composition, CO$_2$ adsorbs at the interface and the degree of interaction depends upon the composition and shape of clusters. The CO$_2$ interaction with Pt$_{13}$/TiO$_2$ is stronger as compared to Cu$_{13}$/TiO$_2$ systems. Although we do not observe any specific trend in adsorption energies, our calculations suggest that the CO$_2$ interaction magnitude would be higher for CuPt clusters than for only pure Cu. As compared to the bond length (1.162 \AA) and angle (180$^\circ$) of CO$_2$ in gas phase, upon its adsorption the C-O bond elongates, and the bond angle changes to 121-134$^\circ$. Taken together results of the CO$_2$ interaction with CuPt/TiO$_2$ systems with various Pt/Cu compositions ($E_{ads}$=-0.50 to -1.2 eV), the charge transfer to an adsorbed CO$_2$ ($\sim$0.7$e$ to 0.9$e$) and the elongation of C-O bond ($\sim$0.2 \AA) and the substantial reduction of $\angle$OCO (55-59$^\circ$), we show that CuPt/TiO$_2$ systems facilitate CO$_2$ activation.

\begin{table}[h]
\caption{Adsorption and activation of CO$_2$ at the interface of Cu$_{(13-n)}$Pt$_n$/TiO$_2$(110), $n=0, ..., 13$, in the lowest-energy configurations. Given are adsorption site, adsorption energy $E_{ads}$ (eV), CO bond length d$_{(CO)}$ (\AA), OCO bond angle $\angle$OCO ($^\circ$), distance between C and Cu or Pt (\AA), and net charge ($e$) gained by the adsorbed CO$_2$ according to the Bader decomposition of the charge density (method details in SI). The d$_{(CO)}$ and $\angle$OCO of the gas phase CO$_2$ are respectively 1.162 \AA \ and 180$^\circ$.}
\label{Table1}
\begin{tabular*}{\textwidth}{@{\extracolsep\fill}lcccccc}

\toprule

Pt$_n$ & Adsorption  & $E_{ads}$ & d$_{(CO)}$ & $\angle$OCO & d$_{(C-Cu)}$ & Charge gained  \\
 & site &  &  &  & d$_{(C-Pt)}$ & by CO$_2$ \\
\midrule
0  & Cu $\&$ O$_{vac}$ & -0.42 & 1.266, 1.321 & 121.4 & 1.940 & 1.00\\
1  & Cu $\&$ O$_{vac}$ & -0.62 & 1.212, 1.370 & 122.9 & 1.921 & 0.87\\
2  & Pt $\&$ O$_{vac}$ & -1.24 & 1.271, 1.306 & 121.3 & 2.006 & 0.95\\
3  & Pt $\&$ O$_{vac}$ & -0.71 & 1.218, 1.320 & 127.7 & 2.028 & 0.78\\
4  & Pt $\&$ O$_{vac}$ & -0.67 & 1.217, 1.315 & 128.3 & 2.020 & 0.75\\
5  & Pt $\&$ O$_{vac}$ & -0.95 & 1.253, 1.287 & 125.5 & 1.997 & 0.79\\

6  & Pt $\&$ O$_{vac}$ & -0.80 & 1.216, 1.352 & 124.1 & 2.021 & 0.80\\
7  & Pt $\&$ $5f$ Ti   & -0.86 & 1.246, 1.251 & 134.6 & 1.977 & 0.54\\
8  & Pt $\&$ O$_{vac}$ & -0.94 & 1.216, 1.364 & 123.0 & 2.020 & 0.82\\
9  & Pt $\&$ O$_{vac}$ & -1.15 & 1.232, 1.358 & 122.5 & 1.987 & 0.96\\
10 & Pt $\&$ $5f$ Ti   & -0.48 & 1.309, 1.217 & 128.2 & 2.016 & 0.64\\

11 & Pt $\&$ O$_{vac}$ & -0.67 & 1.350, 1.215 & 124.6 & 2.017 & 0.79\\
12 & Pt $\&$ $5f$ Ti   & -1.03 & 1.240, 1.265 & 133.6 & 1.961 & 0.53\\
13 & Pt $\&$ $5f$ Ti   & -1.01 & 1.242, 1.266 & 134.4 & 1.951 & 0.58\\
\botrule
\end{tabular*}
\end{table}

Fig.4 and others in Fig.S5 (SI, S2.4) present the lowest-minima structures of CO$_2$@Cu$_{(13-n)}$Pt$_n$/TiO$_2$ configurations. CO$_2$ adsorbs at the interface of Cu$_{13}$/TiO$_2$ such that C attaches to a Cu atom and O$_{(CO_2)}$ atom fills out the O vacancy site (Fig.4a). For Cu$_{(13-n)}$Pt$_n$ with $n \neq 7, 10, 12, 13$ (Pt$_n$ $>$ 1), the CO$_2$ adsorbs at the interface with an O$_{(CO_2)}$ atom occupying O vacancy site and C bonded to Pt, thus preferring a Pt over Cu site (Fig.4b-e). For $n=13$, CO$_2$ adsorbs in a way that C is bonded to Pt and a O$_{(CO_2)}$ to $5f$ Ti (Fig.4f). The CO$_2$ adsorption geometries in the case of $n=7,10,12$ are like that of $n=13$. These results indicate that the CuPt/TiO$_2$ interface serves as active sites for CO$_2$ adsorption, particularly highlighting the importance of interfacial Cu/Pt and surface Ti atoms (near O vacancy). 

From the electron density difference ($\Delta\rho$) plot (Fig.4g), it is seen that the resulting electron density is re-distributed in the bonding regions of *CO$_2$ atoms (C, O1, O2) and interfacial atoms (Pt/Ti). Thus, the favorable interaction between CO$_2$ and Cu$_4$Pt$_9$/TiO$_2$ is mediated by the interface. The charge re-distribution, upon CO$_2$ interaction with Cu$_4$Pt$_9$/TiO$_2$, suggests that the charge transfer takes place between two sub-systems. There is charge accumulation around O$_{(CO_2)}$ atoms and charge depletion around a C atom and surface Ti atoms. Some charge addition to the *CO$_2$ anti-bonding state can trigger the C-O bond elongation. Indeed, based on the partial charges analysis, the net charge transfer of 0.96e takes place from Cu$_4$Pt$_9$/TiO$_2$ to an adsorbed CO$_2$ (Table 1), inducing the C-O bond elongation and thereby facilitating CO$_2$ activation.

\begin{figure}[h]
\centering
\includegraphics[width=0.9\textwidth]{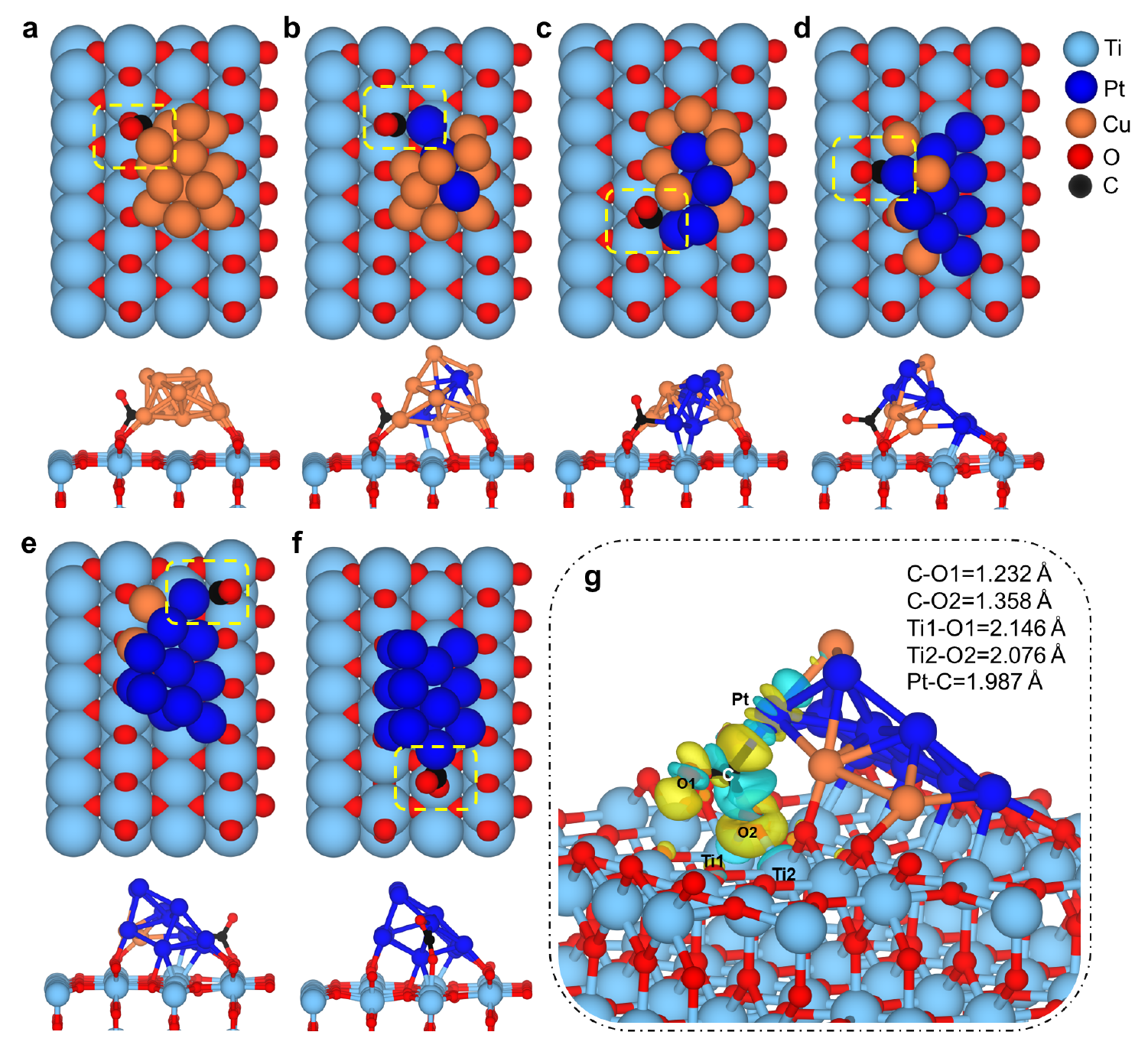}
\caption{Schematic representation of DFT-optimized structures of adsorbed CO$_2$ (dashed yellow boxes) on Cu$_{(13-n)}$Pt$_n$/TiO$_2$(110) for $n$ equal to a) 0, b) 3, c) 5, d) 9, e) 11, and f) 13 (top and side views). In (a-e), an O atom of *CO$_2$ fills out a surface O vacancy, and in (f) an *CO$_2$ O bonds to 5-fold coordinated Ti atom. (g) shows the electron density difference for a configuration (d). The yellow and cyan regions represent the positive and negative iso-surfaces with electron densities of +0.005 $e/a_0^3$ and -0.005 $e$/$a_0^3$ ($a_0$ is Bohr radius), respectively.}\label{fig4}
\end{figure}

Hereinafter, we present the results for structure and energetics properties of Cu$_4$Pt$_9$/TiO$_2$ model system when interacting with the reactants (CO$_2$/H$_2$O) and key intermediates of CO$_2$ reduction. The model selection is partly based on the stronger CO$_2$ adsorption at interfacial sites of Cu$_4$Pt$_9$/TiO$_2$ and partly on the higher Pt concentration than Cu (prevent a CuPt cluster from oxidizing).

\begin{figure}[h]
\centering
\includegraphics[width=0.9\textwidth]{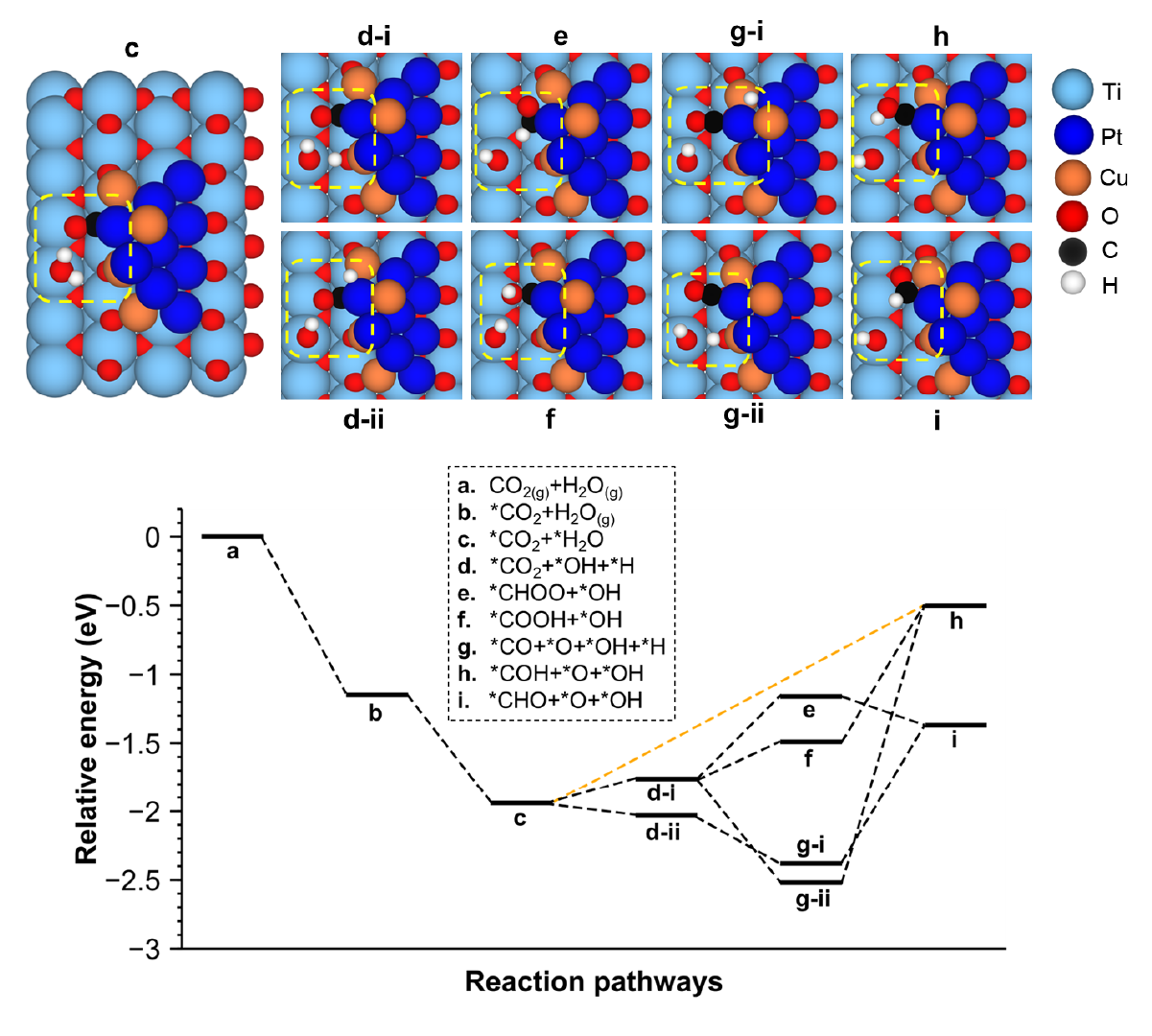}
\caption{Energy profile for elementary reaction steps involving the reactants (CO$_{2(g)}$, H$_2$O$_{(g)}$), CO$_2$ activation and catalytic CO$_2$ conversion to intermediates: (a) CO$_{2(g)}$+H$_2$O$_{(g)}$, (b) *CO$_2$+H$_2$O$_{(g)}$, (c) *CO$_2$+H$_2$O*, (d) *CO$_2$+*OH+*H where *H is adsorbed at a bridging O atom (d-i) and at a Cu-Pt site (d-ii), (e) *CHOO+*OH, (f) *COOH+*OH, (g) *CO+*O$^{ad}_{vac}$+*OH+*H where *H is adsorbed at a Cu-Pt site (g-i) and at a bridging O atom (g-ii), (h) *COH+*O$^{ad}_{vac}$+*OH, and (i) *CHO+*O$^{ad}_{vac}$+*OH. Here, *O$^{ad}_{vac}$ (labelled as *O in inset) refers the adsorbed oxygen atom filling the surface O vacancy site. The top panel shows top views of the atomic configurations (side views are provided in Fig.S6). Dashed yellow boxes highlight the adsorbed species with energetically preferred sites.}\label{fig5}
\end{figure}

\subsection{Energetics for CO$_2$ dissociation vs hydrogenation}\label{subsec2}
Based on the DFT-calculated energetics, the *CO formation from *CO$_2$ dissociation is energetically favorable ($\Delta$E=-0.62 eV). In contrast, the CO$_2$ hydrogenation via *CO$_2$+*H$ \rightarrow$ *CHOO+* step is endothermic ($\Delta$E=+0.60 eV). Similarly, the *CO$_2$+*H$ \rightarrow $*COOH+* is endothermic ($\Delta$E=+0.27 eV). Although these steps require some energy, the reactions might be activated in normal reaction conditions. Calculated energetics provide some indication that while *CO formation is a favorable process (*CO hydrogenation can proceed to yield *CHO formation, being more thermodynamically favorable than *COH formation.\cite{Liu_Xiao}), the *CO$_2$ hydrogenation might also feasible under external driving force.

\subsection{Structures and energetics of intermediates}\label{subsec2}
In Fig.5, we provide the energy profile for reaction between CO$_2$ and H$_2$O (reactants) and some key intermediates adsorbed on Cu$_4$Pt$_9$/TiO$_2$. The relative energy for all adsorbed species (with/without co-adsorption) is given in reference to the sum of energies of gas-phase CO$_{2(g)}$ and H$_2$O$_{(g)}$, and of Cu$_4$Pt$_9$/TiO$_2$. The reaction energy ($\Delta$E) is the difference in energy between initial and final states. The adsorption of H$_2$O in the presence of *CO$_2$ is favorable ($\Delta$E=-0.79 eV). The H$_2$O adsorbs at a Ti site with H being closer to a bridging O atom. The TiO$_2$(110) is well studied for H$_2$O activation.\cite{Bikondoa_Pang, Brookes_Muryn, Wang_Wang, Wen_Andrade, Yuan_Zhu} The CO$_2$ adsorption is energetically favorable ($\Delta$E=-1.15 eV). Based on CO$_2$ activation, we are mainly interested in the chemical activity at/near the interface.

Once *CHOO is formed via hydrogenation step, then it can undergo self-dissociation to *CHO and *O (Fig.5) with the gain of energy ($\Delta$E=-0.21eV). As an alternative pathway, the *CO$_2$ dissociation can occur with *CO formation. The *CO$_2$ dissociation (in co-adsorption with *H at a bridging O atom) is more exothermic ($\Delta$E=-0.62 eV) than from *CO$_2$ dissociation (with *H at a Cu-Pt site) ($\Delta$E=-0.35 eV).

Not only of *CO$_2$ and *CO, but also the stabilities of other intermediates such as *CH, *CH$_2$, *CH$_3$ are important for the selectivity for CH$_{4(g)}$ over CO$_{(g)}$. Now we examine the structure and energetics of *CH$_2$OH (which can be formed via reaction between *H and *CH$_2$O, which can be formed either by *CH$_2$OO dissociation \cite{Kattel_Yan_b} or by a *H+*CHO step.\cite{Kattel_Yan_b, Tang_Zhang} The CH$_2$OH adsorbs at interfacial sites where C attaches to Pt (2.034 \AA) and OH functional group occupies an O-vacancy site (Fig.6a). The dissociation step *CH$_2$OH$+* \rightarrow$ *CH$_2$+*OH is energetically favorable ($\Delta$E=-0.65 eV) and is better than Pt/TiO$_2$ ($\Delta$E=-0.12 eV,\cite{Kattel_Yan_b}) thus indicating that *CH$_2$ formation is thermodynamically feasible and *CH$_2$ is stabilized by interfacial Cu/Pt atoms. These atoms can serve as active sites for *CH$_2$ formation and possibly its further reactions. Since the Pt-H distance (H of *OH) is only 2.322 \AA, the Pt site can facilitate the reaction between *CH$_2$ and *OH, to form *CH$_3$ and *O$^{ad}_{vac}$. For feasibility of this process, however, the *OH bond scission and an additional C-H bond formation should take place simultaneously. The resulting *CH$_3$ then can react with *H to form *CH$_4$ or CH$_{4(g)}$ (Fig.S7). The *CH$_3$+*H$ \rightarrow *+*+ $CH$_{4(g)}$ step can be feasible at room temperature. We note that *CH$_3$ formation route could also proceed with the reaction between *H and *CH$_2$, which is stabilized at the interfacial Cu and Pt atoms, and that *CH$_2$ formation from the reaction between *H and *CH intermediates. Indeed, the formation of *CH from dissociation of *CHOH, which is also activated at the interface, is energetically much favorable ($\Delta$E=-1.28 eV) (Fig.6b).

\begin{figure}[h]
\centering
\includegraphics[width=0.9\textwidth]{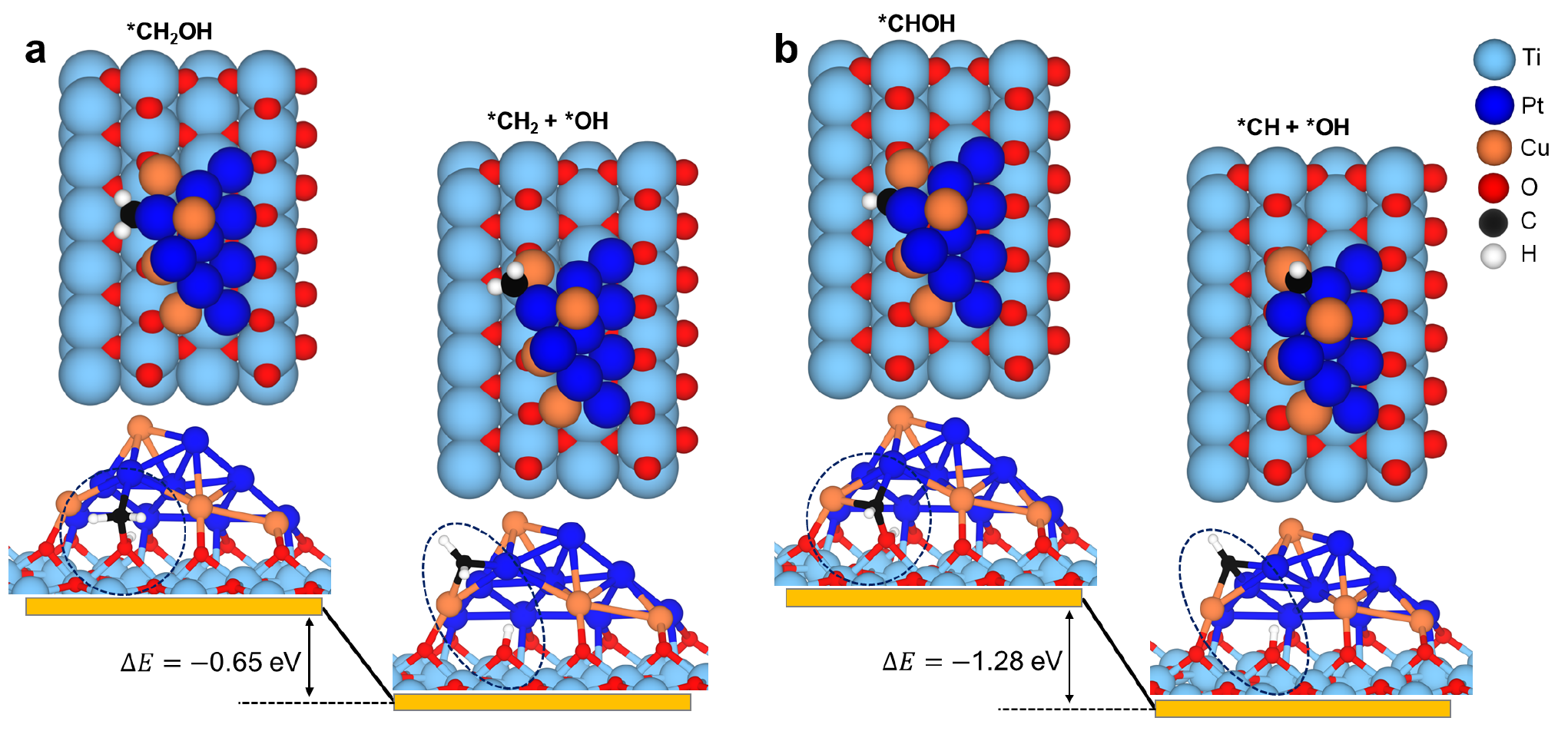}
\caption{Reaction energetics of a) dissociation of *CH$_2$OH into *CH$_2$ and *OH and b) *CHOH into *CH and *OH on Cu$_4$Pt$_9$/TiO$_2$. In the dissociated phase of CH$_2$OH, CH$_2$ adsorbs at a bridge site of interfacial Cu and Pt atoms with bonding of C to Cu-Pt atoms and OH fills O vacancy site with bonding of *OH oxygen with under-coordinated Ti atoms (compared to five-fold coordinated Ti). In the CHOH dissociated phase, CH adsorbs at three-fold site of interfacial Cu and two Pt atoms, whereas OH occupies the O vacancy site. Both reaction steps are energetically favorable with negative $\Delta$E. The top and bottom panels represent respectively the top and side views (the latter slightly rotated for better visibility).}\label{fig6}
\end{figure}

\subsection{Experimental rationalization of CO$_2$ photoconversion}\label{subsec2}
We characterize the morphology and composition of reduced P25 samples by scanning electron microscopy (Fig.7a) and energy-dispersive X-ray spectroscopy (Fig.S8), respectively. The reduced P25 is irregular, with a particle size $<$ 50 nm. We observe some high-contrast features after photodeposition reaction of Pt and Cu. The Pt detected is similar with input Pt amount. In contrast, the Cu detected is significantly less than the input Cu amount with 0.14Pt-0.6Cu and 1.26Pt-0.4Cu requiring $\sim$2.5\% and 1.25\% Cu, respectively. This incomplete photodeposition of Cu compared to Pt might be due to the lower reduction potential of Cu$^{2+}$ than Pt$^{2+}$ and the redox nature of Cu.

\begin{figure}[h]
\centering
\includegraphics[width=0.9\textwidth]{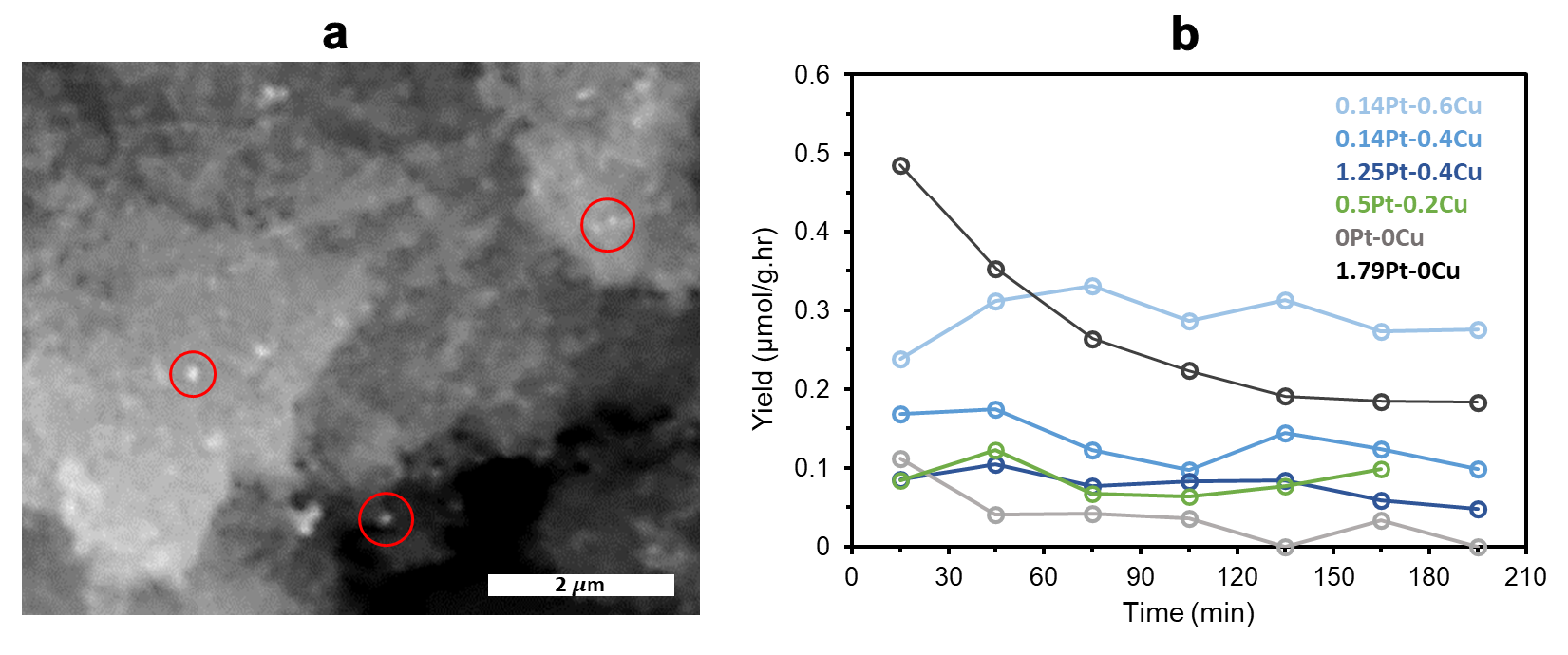}
\caption{Morphology and photocatalytic activity of CuPt/P25 samples. a) Scanning electron microscopy (SEM) micrograph of 1.25Pt-0.4Cu (in mol\%) with red circles highlighting some areas of high contrast likely associated with high Pt/Cu contents. b) CH$_4$ production of reduced P25 decorated with CuPt nanoparticles with various compositions (in mol\%) of Pt and Cu added by photodeposition.}\label{fig7}
\end{figure}

We perform photocatalytic reactor measurements under a flow of humidified CO$_2$ and Ar, illuminated by an AM1.5 filtered Xe lamp focused to 5 sun intensity. In Fig.7b, we show the CH$_4$ yield over time of photocatalysis reactions on CuPt/P25 samples with various CuPt compositions. The sustained CH$_4$ production is the highest for a 0.14Pt-0.6Cu sample (with the photodeposition of 0.14 mol\% Pt and 2.5 mol\% Cu onto the reduced P25), in agreement with an earlier experiment.\cite{Sorcar_Hwang_b} The low Pt and high Cu composition yields higher CH$_4$. The CH$_4$ yield was higher initially for 1.79Pt-0Cu samples, but over time it became deactivated.  Reduced Cu-P25/Pt-P25 or reduced P25 (pure) samples produce significantly less CH$_4$, suggesting that the higher CH$_4$ yield is the consequence of the synergistic effect of Cu, Pt, and P25. The CO$_2$ photoconversion activity decreases substantially as both Pt and Cu loadings are increased, attributed to the growth of large-sized CuPt particles and the reduction of interfacial density of Cu/Pt-TiO$_2$ sites, suggesting the TiO$_2$-Pt/Cu interface is highly preferable for achieving good photocatalytic conversion of CO$_2$ to hydrocarbon.

\section{Discussion}\label{sec3}
Compared to oxide supports \cite{Acharya_Camillone, Lin_Yoon} and metal surfaces,\cite{Allegretti_O’Brien, Schaub_Thostrup} CO$_2$ is chemically well-activated by metal/oxide interfaces. From our DFT simulations, the defect-laden TiO$_2$(110) does not facilitate CO$_2$ activation ($E_{ads}\sim$ -0.2 eV), with an onset desorption temperature $\sim$175 K.\cite{Allegretti_O’Brien} Since TiO$_2$ itself does not stabilize CO$_2$,\cite{Allegretti_O’Brien} CO$_2$ reduction is not feasible. On well-defined metal surfaces, CO$_2$ adsorption is also weak, e.g. $E_{ads}=\sim$0.21-0.28 eV for Cu \cite{Muttaqien_Hamamoto} and $E_{ads}=\sim$0.03 eV for Pt \cite{Liu_Sun} and thus limits to trivial charge transfer between CO$_2$ and Cu/Pt catalysts. Due to weak interaction with ($E_{ads}$=-0.29 eV),\cite{Liu_Sun} it can also desorb before its reaction with H.\cite{Kattel_Yan_b} When supported using TiO$_2$, metal/oxide systems can enhance CO$_2$ adsorption. While Pt$_{25}$/TiO$_2$(110) binds CO$_2$ with $E_{ads}$=-0.61 eV,\cite{Kattel_Yan_b} Cu$_{10}$/TiO$_2$(110) does with $E_{ads}$=-0.65 eV.\cite{Barlocco_Maleki} As compared to bare TiO$_2$ or Cu/TiO$_2$, the CuPt/TiO$_2$ can be considered as a better catalyst owing to: 1) the higher *CO$_2$ stability at the interface, and 2) the interfacial Cu and Pt atoms providing sites for nucleophilic and electrophilic adsorbates.

For the catalytic conversion of CO$_2$ into hydrocarbons, the atomic hydrogens (from H$_2$O splitting process) are required to be present on the catalyst surfaces so that they can react directly with either *CO$_2$ reactant or *CO/*C to form CH-containing intermediates. The CuPt/TiO$_2$ interface provides active sites for formation of intermediates: *CO, *CHOO, *CHO, *CH$_2$OH, *CH, *CH$_2$, and *CH$_3$, thus highlighting the importance of interfacial sites in the chemical activity of CuPt/TiO$_2$ systems towards the CO$_2$ conversion to hydrocarbons.

We propose the reaction pathways as follows: CO$_2$ adsorbs at the CuPt/TiO$_2$ interface with CO$_2$ oxygen at a O vacancy site, and H$_2$O adsorbs at a $5f$ Ti site of the reduced P25, thus facilitating the formation of *H* and *CO at the interfacial Pt sites and leaving behind a surface O$^{2-}$. Then, they can react to form intermediates required for CH$_4$ formation pathway. Finally, the O-vacancies generation is assisted by the redox activity of CuPt nanoparticles with some fraction of Cu on their surface.

Although MLIP/DFT simulations and experimental observations cannot be directly matched, our studies together can provide meaningful insights into CuPt/TiO$_2$ photocatalytic systems. Their excellent chemical activity can be correlated with the thermodynamically favorable, moderate-to-strong interaction ($E_{ads}$=-0.67 to -1.24 eV) with CO$_2$ and the net charge gained by CO$_2$ ($\sim0.7-1.0e$). The synergistic effect of the surface TiO$_2$, Cu, and Pt atoms at the interface results in stronger CO$_2$ adsorption with $E_{ads}$=-0.67 to -1.24 eV), charge transfer of $\sim0.7-1.0e$ to *CO$_2$, CO$_2$ bond elongation, and change in $\angle$OCO to 121-134$^\circ$, indicating CO$_2$ activation. For a Cu$_{11}$Pt$_2$ cluster, approximately equivalent to the experimental composition of 0.14Pt-0.6Cu, the CO$_2$ adsorbs at both O vacancy and Pt sites. It is notable that CO$_2$ strongly interacts with Cu$_{11}$Pt$_2$/TiO$_2$ ($E_{ads}$ $\sim$ -1.2 eV) together with C-O bond elongation of $\sim$0.2 \AA, $\angle$OCO change to $\sim$121$^\circ$, and $\sim0.9e$ gained by *CO$_2$, thus promoting CO$_2$ activation. The preference of *CO$_2$ at Pt facilitates reaction with H* to lead CH$_4$ formation pathway rather than CO desorption ($E_{ads}$=-1.78 eV). The optimal Pt-Cu composition leads to better activity than only Cu. For higher Pt concentration, the selectivity for CH$_4$ was observed but with lower yield. The CO$_2$ activation at the CuPt/TiO$_2$ interface, and the stability of *CO$_2$ and *CO at the interfacial Pt atoms can support the observed CH$_4$ selectivity.

\section{Conclusion}\label{sec4}
In summary, using unbiased MLIP+BHMC together with \textit{ab-initio} DFT approaches, we demonstrate that Cu$_{(13-n)}$Pt$_n$/TiO$_2$ systems strongly facilitate CO$_2$ activation at the interfacial sites. The process is facilitated by charge transfer and direct interaction. The calculated CO$_2$ adsorption energies are from -0.5 eV to -1.2 eV, depending upon Cu/Pt ratio. The VB-maximum frontier states spatially localized at interfacial atoms, suggesting these atoms would be active sites for chemical reactivity.

The interface plays a key role in influencing the chemical activity of CuPt/TiO$_2$ systems. The *CO$_2+* \rightarrow$ *CO+*O step, where oxygen fills the vacancy site and CO adsorbs at an interfacial Pt site, is energetically favorable. The other elementary steps can result in *CHOO, *CHO, *CH$_2$OH, *CHOH, *CH, and *CH$_2$ intermediates. The interface, with optimal CuPt composition, bolsters the stability of *CO$_2$ and *CO, suggesting higher activity and CH$_4$ selectivity. Our simulations qualitatively agree with the experimental observation of photocatalytic conversion of CO$_2$ to CH$_4$. The maximum CH$_4$ yield of $\sim$0.3 $\mu$mol$^{-1}g^{-1}hr^{-1}$ for CuPt composition with 0.14 mol\% Pt and 2.5 mol\% Cu, can be correlated with CO$_2$ activation and CH-intermediates formation. The photocatalytic performance degrades with the decrease of density of interfacial sites, suggesting the necessity of optimal surface area of interfacial regions for selective CO$_2$ conversion to hydrocarbons.

Comprehensive studies of several intermediate structures together with energetics and kinetics of the possible reaction pathways could shed more light on reaction mechanisms. Nonetheless, exploring PES of main reactants (H$_2$O/CO$_2$) and several CH-containing intermediates with activation energy, vibrational entropy, zero-point energy are computationally intensive, and is beyond the scope of the current work. We hope our studies encourage pursuing further study on CuPt/P25 or similar systems and motivate for developing the sophisticated MLIP potential for simulations of these photocatalytic systems with the reactants and intermediates.

\section{Methods}\label{sec5}
\subsection{\textit{Ab-initio} calculations}\label{subsec5}
We performed the \textit{ab-initio} DFT calculations using VASP \cite{Kresse_Furthmüller} and QE \cite{Giannozzi_Andreussi} on AWS EC2 computing platform. The MLIP, as shall be discussed, was trained on VASP generated datasets and a few DFT calculations were performed and validated using QE simulations. We used the plane-wave basis set and pseudopotential approaches. For the exchange-correlation of electrons, we used the generalized-gradient approximation (GGA) in the form of Perdew–Burke-Ernzerhof (PBE) functional.\cite{Perdew_Burke} We used the projector-augmented wave (PAW) pseudopotential method \cite{Blochl_1994} for describing electron-ion interactions. Here, we chose the rutile TiO$_2$(110) since it is the most stable among its various surfaces and extensively studied model system.\cite{Pang_Lindsay} We constructed the 13-atom sub-nanometer-sized CuPt nanoclusters, denoted by Cu$_{(13-n)}$Pt$_n$. The motivation for the choice of magic-numbered clusters partly arises from several earlier computational studies, e.g.\cite{Rawal_Le, Imaoka_Kitazawa, Almeida_Chagoya, Hong_Rahman, Zhang_Wang} and partly come from the computationally tractable combinatorial problem. Further details on DFT calculations can be found in supporting information (SI) section S3.1, on energetics calculations in section S2.2 and Bader charge analysis in section S2.3.

\subsection{Machine learning interatomic potential}\label{subsec5}
To solve a combinatorial problem relating to complexities of structures involving five different chemical elements, we built a machine learning model based on Allegro – a deep equivariant neural network architecture.\cite{Mazheika_Wang} We use a 6 \AA \ radial cutoff and 2 interaction blocks for the Allegro-based MLIP. A polynomial envelope with cutoff of $p=6$ and eight trainable Bessel functions are used for a basis expansion of radial distances. We restrict the maximum irreducible representation of SO(3) for internal rotational features (i.e., the maximum order of geometric tensor embeddings transforming like type-l spherical harmonics) to $lmax=2$. For each $l$, we allowed for both even and odd parity features. The two-body embeddings consist of 3 layers of dimensions (32, 64, 128). Within each interaction block, we use latent MLIPs consisting of 3 layers of dimensions (128, 128, 128). As nonlinear activations, sigmoid linear unit (SiLU) functions are used throughout the entire architecture. MLIP optimizers, learning rates and numerical precisions are discussed in the SI section S3.4.

We employ an iterative training approach following the creation of an initial potential. Structures generated from basin-hopping Monte Carlo (BHMC) simulations,\cite{Sumaria_Nguyen, Sumaria_Sautet} accelerated by the trained MLIP, go through a workflow for identifying unique configurations (utilizing uniform manifold approximation and projection (UMAP) and K-means clustering). Single-point DFT calculations on configurations exhibiting high errors relative to MLIP predictions are incorporated into the training set. Iterations persist until convergence in terms of energy ($<$ 1 meV/atom) and force errors ($<$0.1 eV/\AA) is achieved. Additional details about the method are provided in the SI (see section S3.5 and S3.6).

\subsection{Sampling unbiased training}\label{subsec5}
The MLIP training dataset, previously described, encompasses atom coordinates, energy, and forces of CO$_2$@Cu$_{(13-n)}$Pt$_n$/TiO$_2$ systems. The (6x2)-TiO$_2$(110) substrate, featuring a single oxygen vacancy, interacts with Cu$_{(13-n)}$Pt$_n$ nanoclusters ($n=0$ to 13), and CO$_2$ is positioned at various adsorption sites. Overall, the training data includes $\sim1.9\times10^6$ bulk atoms (TiO$_2$ substrate) and $\sim0.73\times10^6$ surface atoms (consisting of 4,690 C and O atoms from CO$_2$ and $\sim0.12\times10^6$ Pt/Cu atoms). As a result, the MLIP training inherently favors TiO$_2$ atoms to reduce overall prediction errors for the predicted forces, sacrificing for surface Ti and O atoms as well as the atoms comprising the nanoparticles and the adsorbate, which are crucial for understanding the surface science and catalytic activity.

This bias can be quantified by defining a gaussian density function (GDF) for the encoding of an atomic environment that describes the radial and angular distribution (Symmetry functions \textbf{G}) of the neighboring atoms within a certain cutoff radius (see SI section S3.5). For an arbitrary \textbf{G} for an atomic species in the entire space, we define the Gaussian Density function $\rho(\textbf{G}$) as: 

\begin{equation}
\rho(\mathbf{G})=\frac{1}{M} \sum_{j=1}^M \exp \left(-\frac{1}{2 \sigma^2} \frac{\left|\mathbf{G}-\mathbf{G}_j\right|^2}{D}\right)
\label{eq:GDF}
\end{equation}

where $\sigma$ is the Gaussian width and $D$ is the dimension of the symmetry function vector, $M$ is the total number of atoms in the entire dataset. The $\rho(\textbf{G})$ ranges between 0 and 1, where $\rho(\textbf{G})\sim$0 means scarce training points whereas $\rho(\textbf{G})\sim$1 represents abundant training points.

Addressing biases arising from the redundancy of atomic environments in the training set, we also acknowledge non-uniformity in training on atomic forces. The existing loss functions for MLIP training treat the absolute error in forces as constant, irrespective of force magnitude, leading to higher relative force errors for smaller values. To overcome the identified issues, we propose the use of an alternative weighted loss function in the MLIP model ensuring improved and uniform training for the \textit{ab-initio} DFT data. The modified loss function takes the following form: 

\begin{equation}
\begin{aligned}
\Gamma= \frac{1}{N} \sum_{i=1}^N\left(\frac{E_i^{\mathrm{DFT}}-E_i^{\mathrm{NNP}}}{n_i}\right)^2+ \frac{\mu}{3 M} \sum_{j=1}^M \left[\Theta\left[\frac{1}{\rho\left(\mathbf{G}_j\right)}\right] + A e^{-|a\mathbf{F}_j^{\mathrm{DFT}}|} \right] 
 \left|\mathbf{F}_j^{\mathrm{DFT}}-\mathbf{F}_j^{\mathrm{NNP}}\right|^4
\end{aligned}
\label{eq:loss_func_new}
\end{equation}
\begin{equation}
\Theta(x)=\frac{B x}{1+\mathrm{e}^{-b x+c}}
\label{eq:mono}
\end{equation}

where B is a scaling factor used to tune the impact of low-force data on the loss function (chosen to be 10 here), $\mu$ determines the relative weight of force error with respect to the energy error in the loss function ($\mu=1$ here), $\Theta$ is a monotonically increasing function (modified sigmoid function), A is a normalizing constant (that makes the average of $\Theta$ to be 1), and b and c are parameters that are ﬁne-tuned for the balanced training (chosen to be 150 and 1.0, respectively).

\subsection{Experimental details}\label{subsec5}

\subsubsection{Synthesis of CuPt nanoparticles decorated with P25}\label{subsubsec5}
The CuPt nanoparticles decorated with P25 were synthesized using methods as reported by Sorcar et al.\cite{Sorcar_Hwang_b} Before synthesis of CuPt nanoparticles, the reduced P25 was initially prepared by mixing 200 mg of P25 with 30 mg of NaBH$_4$ in a mortar and pestle. The powder was then heated in an inert atmosphere (Ar flow) at 350$^\circ$C for about an hour. Upon cooling, the P25 was purified by washing in water and ethanol and then the sample was centrifuged. This was repeated at least five times. The resulting powder was dried overnight in a vacuum oven at 100$^\circ$C.

The platinum was deposited onto the reduced P25 via photo-deposition. 40 mg of the P25 was initially dispersed in 10 mL of the 4:1 H$_2$O:CH$_3$OH solution. H$_2$PtCl$_4$ solution was then added to the dispersion and allowed to stir for one hour. The dispersion was then irradiated by AM1.5 light by a 300 W Xe lamp at 1 sun intensity for two hours. The Pt-P25 was collected by centrifuging. Then it was washed with H$_2$O and ethanol mixture three times and dried under vacuum. Copper was subsequently deposited in a similar method using Cu(NO$_3$)$_2$ instead of H$_2$PtCl$_4$. The CuPt-P25 samples were then deposited onto glass frits before loading into the photoreactor. Dispersions of the samples in isopropanol were drop-coated onto Aceglass glass frits with porosity C until about 15 mg was loaded. The frits were then dried under vacuum overnight.

\subsubsection{Experimental setup for measuring photocatalytic activity of CuPt-P25 }\label{subsubsec5}
The catalyst coated glass frit was placed into a custom-built steel photoreactor fitted with a quartz window. Before testing, the reactor was purged by evacuating the reactor and refilling with 1:1 Ar:CO$_2$ ten times. 1.5 sccm of Ar and 1.5 sccm CO$_2$ were bubbled through a water filled gas washing bottle and fed into the reactor from the top, passing through the frit, and exiting the bottom of the reactor. The reactor was then irradiated by a 300 W Xe lamp at one sun intensity through an AM1.5 filter. The product gas composition was analyzed by a gas chromatograph (SRI GC MG5).


\bmhead{Acknowledgements}
We acknowledge Niraj Prasad for his valuable suggestions in setting up of the compute clusters on AWS EC2 platform that allows us to run first-principles DFT calculations and to train machine learning models. We thank Alexander Imbault for reading the manuscript and providing his comments. We would also like to acknowledge George Hathaway and Jun Cao at Hathaway Research International for their invaluable advice to this study’s physical experimental efforts.

\backmatter
\bmhead{Supplementary information}
The supporting information (SI) contains the additional figures (Fig.S1-S9) and descriptions of results and methods. We describe the adsorption of CO$_2$ on a bare rutile TiO$_2$ surface (section S1.1), MLIP training and validation (section S2.1), exploration of potential energy surface (section S2.2), adsorption geometry of CuPt/TiO$_2$ (section S2.3), CO$_2$ adsorption on CuPt/TiO$_2$ (section S2.4), structures of reactants and intermediates and product (section S2.5), experimental measurements (section S2.6), and methods (section S3).   

\bmhead{Competing interests}
We have filed a provisional patent relating to this work.

\bmhead{Author contributions}
V.S. and T.B.R. performed the \textit{ab-initio} DFT calculations and analyzed the data. R.J.B. performed the initial DFT calculations. Y.F.L. synthesized the materials, carried out photocatalysis experiments, and analyzed the experimental data. V.S. performed the MLIP training and validation; and used the modified basin-hopping Monte Carlo algorithm for data generation. D.S., J.V., A.P., and D.P. proposed the application of E(3)-equivariant neural networks and participated in the discussions with the development of MLIP potential. V. S., T.B.R., and Y.F.L. wrote the manuscript in discussions with all co-authors who also provided some contributions in writing this manuscript. T.B.R., Y.F.L., A.P., D.P. conceived the project. M.R. advised during the research process.



\section*{Supplementary material}

# Supporting Information for
# Machine Learning, Density Functional Theory, and Experiments to Understand the Photocatalytic Reduction of $CO_2$ by CuPt/$TiO_2$

Vaidish Sumaria[1,2], Takat B. Rawal[1*], Young Feng Li[1], David Sommer[1],
Jake Vikoren[1], Robert J. Bondi[1], Matthias Rupp[1,3], Amrit Prasad[1],
Deeptanshu Prasad[1]

[1*]Quantum Generative Materials (GenMat), 411 W. Monroe St, Austin, TX 78704, USA.
[2]Caminosoft Technologies Inc., 1197 E Los Angeles Ave St C305, Simi Valley, CA 93065, USA.
[3]Luxembourg Institute of Science and Technology (LIST), Belvaux, Luxembourg.

*Corresponding author(s). E-mail(s): takat.rawal@genmat.xyz;

# S1 Introduction

## S1.1 Adsorption of $CO_2$ on bare $TiO_2$

Using *ab-initio* DFT approach, we firstly examine the $CO_2$ activation on the bare rutile $TiO_2$(110) support without metal nanoclusters. Such examination will allow us to determine how these metal clusters play a role in $CO_2$ activation. We study the adsorption of $CO_2$ on $TiO_2$(110) in presence of O vacancy.

Our calculations indicate that $CO_2$ weakly interacts with $TiO_2$(110) with single O vacancy. For $CO_2$ adsorption at an O vacancy site (Fig.S1a), the adsorption energy is -0.23 eV. The adsorption at a 5-fold ($5f$) coordinated Ti site (Fig.S1b) is further weaker with $E_{ads}$ of -0.13 eV. Although adsorption energy is only decreased by ∼0.1 eV, the $Ti^{4+}$ sites are found to be less active than $Ti^{3+}$. For a $5f$ Ti site, the shortest distance between one of $CO_2$ oxygen and Ti atom is 2.762 Å and the longer C-O bond length is 1.177 Å and ∠OCO is 179.3°. For $CO_2$ adsorption at an O vacancy site (with $Ti^{3+}$ centers prior to adsorption), it is 2.675 Å and the C-O bond is 1.182 Å and ∠OCO is 179.7°. Thus, the bare $TiO_2$(110) support does not trigger $CO_2$ activation without any external perturbation.

Since the $CO_2$ adsorption on the bare $TiO_2$ support is weaker, it is important to introduce metal nanostructure sub-systems (as co-catalysts) with d orbitals. While an earlier experiment [1] suggests that the C-O bond from $CO_2$ rather than the O–H bond from $H_2O$ is the rate-determining step in $CH_4$ formation, another experimental study[2] indeed shows that CuPt/titania is a good photocatalyst system for $CO_2$ conversion to hydrocarbons. Here, we consider a model CuPt/$TiO_2$ system using the 13-atom sub-nanometer-sized CuPt clusters supported on $TiO_2$(110) (We consider (6x2)-$TiO_2$ supercell to accommodate $Cu_{(13-n)}Pt_n$ and to ensure the spurious electrostatic interaction between periodic images of $Cu_{(13-n)}Pt_n$ nanoclusters be prevented). Since earlier studies [3–7] have proven that metal-support interaction affects the chemical activity of the supported clusters, we expect that the interaction between CuPt and $TiO_2$ subsystems and the charge transfer between them might help in $CO_2$ activation and its conversion to intermediates that could participate in reaction pathways towards formation of hydrocarbon products.

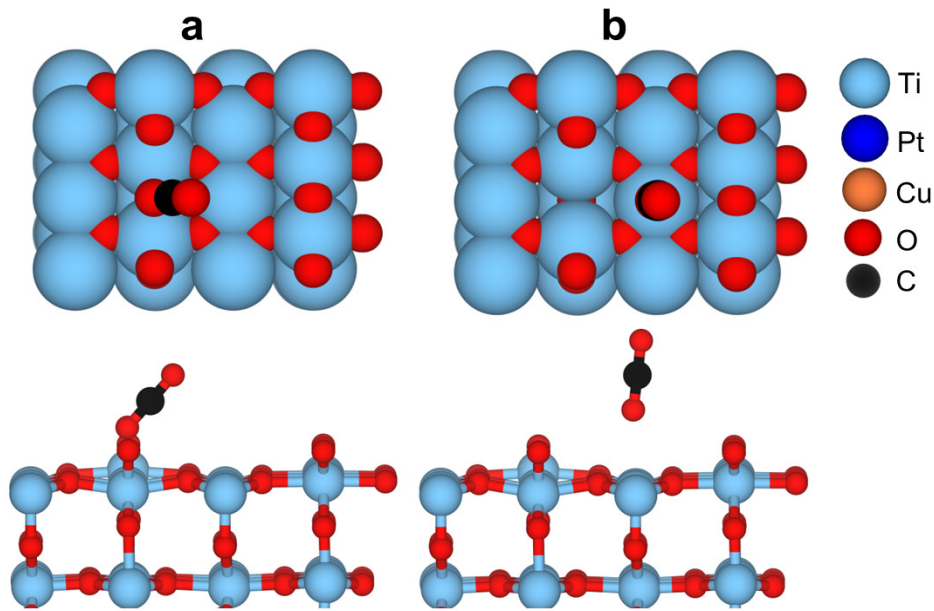

**Fig. S1**: DFT-optimized geometry of a $CO_2$ molecule adsorbed a) at an oxygen vacancy site and b) atop a five-fold coordinated Ti site of the (3×2) rutile $TiO_2(110)$ with a single O vacancy. The top and bottom panels represent respectively the top and side views.

## S2 Results

### S2.1 MLIP training and validation

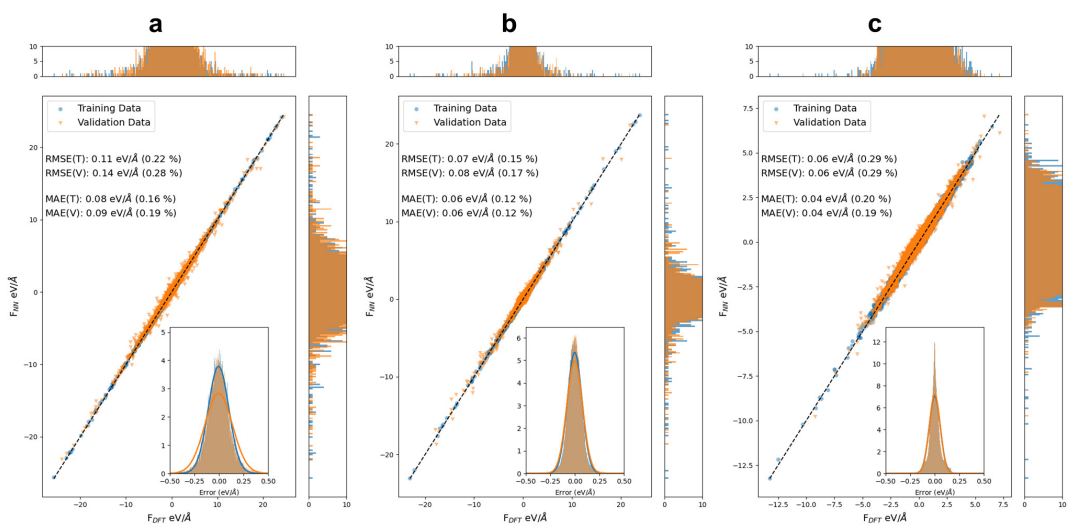

**Fig. S2**: Parity plots showing the decomposed MLIP predicted forces vs the DFT calculated forces for the: a) $CO_2$ adsorbate, b) Cu/Pt/CuPt nanoclusters and c) $TiO_2(110)$ support. For each case, the histograms of the distribution of data points are shown in top and right panels. Insets: The distribution of prediction errors.

In Fig.S2, we show the decomposed force parities for $CO_2$ adsorbate, Cu/Pt nanoclusters, and $TiO_2(110)$ support. To compare the effect of weighting the loss function with the gaussian density function base b weights and absolute force-based weights, we plot the histogram of force data that show errors >0.1 eV/Å. Majority of errors coming from the weighted MLIP distribution lies in lower error region (< 0.25 eV/Å). This effect is particularly prominent for the nanocluster atoms, suggesting that the loss-function weighing has a considerable impact on the overall performance of MLIP. Fig.S3

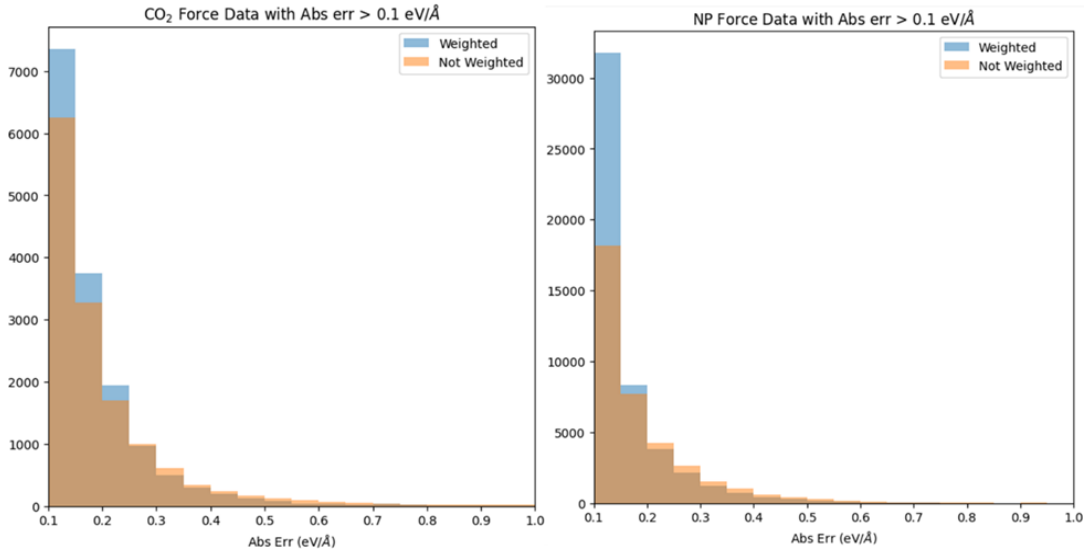

**Fig. S3**: Histogram of force data that show errors > 0.1 eV/Å for the underrepresented data in the entire dataset i.e. the forces on the $CO_2$ adsorbate and nanocluster atoms.

presents the histograms of force data showing the errors >0.1 eV/Å for the underrepresented data, i.e. the forces on atoms of $CO_2$ and nanoclusters, of the entire dataset.

### S2.2 Exploration of potential energy surface

We start with a random configuration of $CO_2$ adsorbed at a Pt atom of the CuPt cluster. In our modified version of BHMC, the custom modifiers allow efficiently to modify shapes and configurations of CuPt, yielding efficient exploration of various adsorption sites for $CO_2$, including the interfacial atoms. As the simulation progresses, the initial random-seed structure is modified using one such modifier and the so-generated structure is relaxed to a local basin using the MLIP. Using the MC criteria, the new structure is either accepted if lower in energy (or if the energy difference is smaller than the hopping temperature) or rejected if significantly higher in energy. In this manner, the simulation efficiently searches for structures within one basin and hops to either a lower energy basin to find a new lower-energy configuration or to a higher-energy one if stuck in some minimum. If no new minima are found after a considerable number of new structures generated (roughly 50) by the algorithm, the simulation is stopped, and an ensemble of low energy configurations is analyzed using DFT calculations. Even with considerably low validation errors of the MLIP, using its predicted energies in a small energy range to compare the stability among structures is not viable. As a result, we choose an arbitrary small energy window of 0.1 eV, such that any structure generated using BHMC that has energy difference <0.1 eV compared to the global minimum predicted is added to the low-energy configurational ensemble. These structures in the ensemble are relaxed using DFT to avoid any small MLIP errors. As shown in Fig.2 (main text), every 25th structure is approximately generated within the global minima search. The extremely large space is apparent from an illustration of PES exploration for a model system, whose configurations are found to be different CuPt-cluster shapes with various arrangement of Pt/Cu atoms together with several possible adsorption sites for $CO_2$.

### S2.3 Adsorption geometry of CuPt on $TiO_2$

In Fig.S4, we present the distribution of Cu-O distances in $Cu_{(13-n)}Pt_n/TiO_2(110)$ systems with $n = 0, ..., 12$. There is a variation of distance between each interfacial Cu atom and $TiO_2$ bridging O atom. For the selected $TiO_2$-supported nanoclusters, the calculated average distances are as follows: Cu-O=1.911 Å for $n = 0$; Cu-O=1.892 Å for $n = 3$; Cu-O=1.917 Å, Pt-Ti=2.870 Å for $n = 7$; Cu-O=1.888 Å and Pt-Ti=2.599 Å for $n = 9$; Cu-O=1.908 Å and Pt-Ti=2.694 Å for $n = 11$; Pt-O=2.060 Å and Pt-Ti=2.725 Å for $n = 13$.

### S2.4 Adsorption of $CO_2$ on CuPt/$TiO_2$

In Fig.S5, we show the selected configurations of $CO_2$ adsorbed on $Cu_{(13-n)}Pt_n/TiO_2$ systems, with $n = 1,2,4,6,8$, and 12. For all configurations, $CO_2$ adsorbs at the interfacial sites. In Fig.S5(a-e), one of

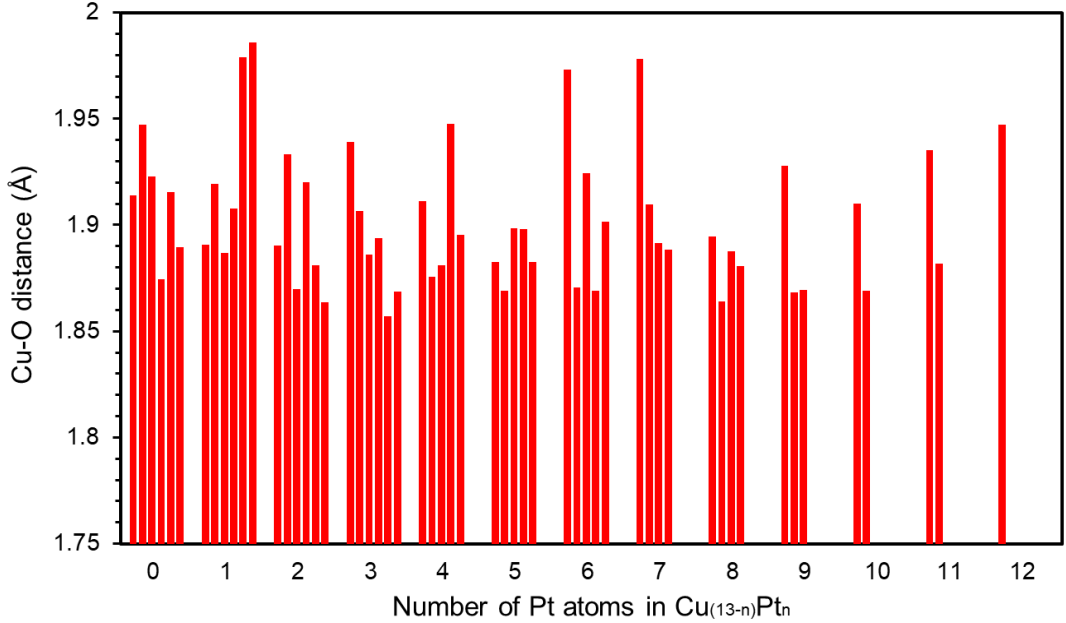

**Fig. S4**: Distribution of distances between surface O and interfacial Cu atoms of $Cu_{(13-n)}Pt_n/TiO_2(110)$ systems, with $n = 0, ... 12$. For $n = 0, 1, 2, 3$, six Cu atoms make bonds with bridging O atoms; five Cu atoms for $n = 4, 5, 6$; four Cu atoms for $n = 7, 8$; three Cu atoms for $n = 9$; two Cu atoms for $n = 10, 11$; only one Cu for $n = 12$.

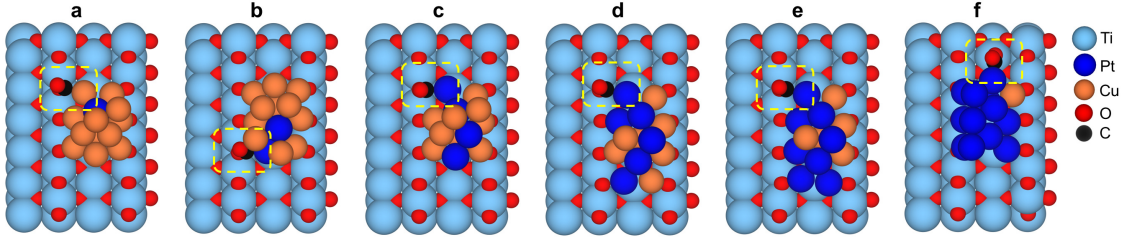

**Fig. S5**: (Top views) Schematic representation of DFT-optimized structures of $CO_2$ adsorbed on $Cu_{(13-n)}Pt_n/TiO_2(110)$ for $n$ equal to: a) 1, b) 2, c) 4, d) 6, e) 8, and f) 12.

O atoms of an adsorbed $CO_2$ fills the O vacancy, and in Fig.S5(f) one of $CO_2$ O atom is bonded to $5f$ Ti atom.

## S2.5 Structures of reactants, intermediates and product

In Fig.S6, we show the DFT-optimized structures of adsorbed reactants ($CO_2$ and $H_2O$) and intermediates which could involve in the reaction pathways towards $CH_4$ formation. Since $CO_2$ preferentially adsorbs at the interfacial sites, we allow interaction of $H_2O$ with an adsorbed $CO_2$ by placing $H_2O$ at a neighboring atop a $5f$ coordinated Ti site. The local interaction of $H_2O$ with $5f$ Ti atom has been realized in earlier experiments.[8] After geometry relaxation, the $H_2O$ adsorbs at the $5f$ Ti site in a configuration (Fig.S6a), where the Ti-O distance is 2.206 Å, the shortest distance between $H_2O$ H atom and bridging O atom is 2.286 Å, and the shortest distance between $H_2O$ H atom and $CO_2$ O atom is 1.845 Å. Once the O-H bond scission takes place, atomic H can transfer to a bridging O atom, which is also bonded to a Cu atom. Now, the *OH still favors a $5f$ Ti site (Fig.S6b), where the distance between an O atom of *OH and a $5f$ Ti site is 1.878 Å, the distance between O of *OH and *H of dissociated $H_2O$ is 1.731 Å, and the shortest distance between *H (from dissociated $H_2O$) and an interfacial Cu atom is 2.373 Å (It is likely that Cu atoms may facilitate the migration of atomic H to Pt sites). The adsorption of H to a bridging O atom slightly impacts the adsorption geometry of *$CO_2$ with C-Pt distance of 1.983 Å. Note that the adsorption of atomic H (from dissociated *$H_2O$) either at a bridging O atom (bonded to Cu) or bridging O atom (bonded to Pt/C atoms) results in the same stability of the adsorbed systems. The shortest distance between *H (of dissociated *$H_2O$) and C atom is 3.027 Å. The

4

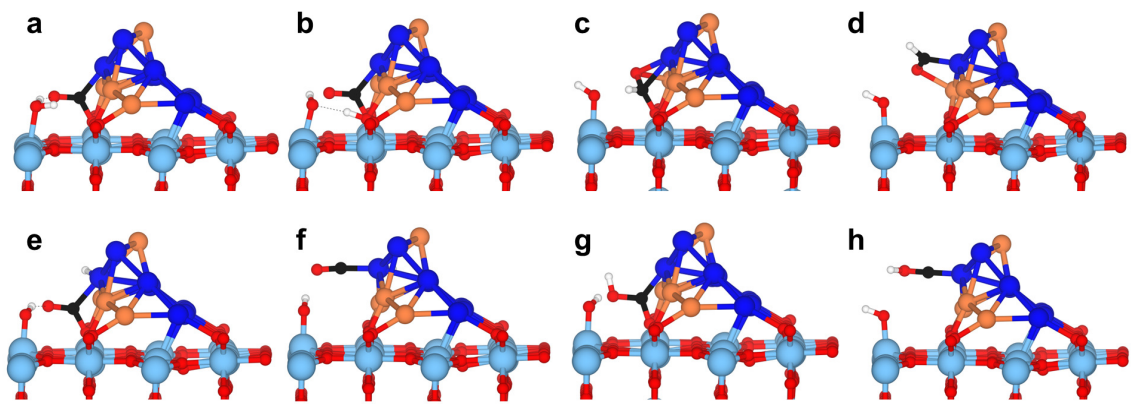

**Fig. S6**: (Side views) DFT models of adsorbed reactants and intermediates on a $Cu_4Pt_9/TiO_2$ model system: a) $*CO_2+*H_2O$, b) $*CO_2+*OH+*H$ (where *H is at bridging O atom), c) $*CHOO+*OH$, d) $*CHO*+*O_{vac}^{ad}+*OH$, e) $*CO_2+*OH+*H$ (where *H is adsorbed at a Pt site), f) $*CO+*O_{vac}^{ad}+*OH+*H$, g) $*COOH+*O_{vac}^{ad}+*H$, and h) $*COH+*O_{vac}^{ad}+*OH$. Here, $O_{vac}^{ad}$ represents the $CO_2$ oxygen occupies the O vacancy site. For hydrogen atom, the color code is white, and for other atoms, the color code is same as in Fig.S5.

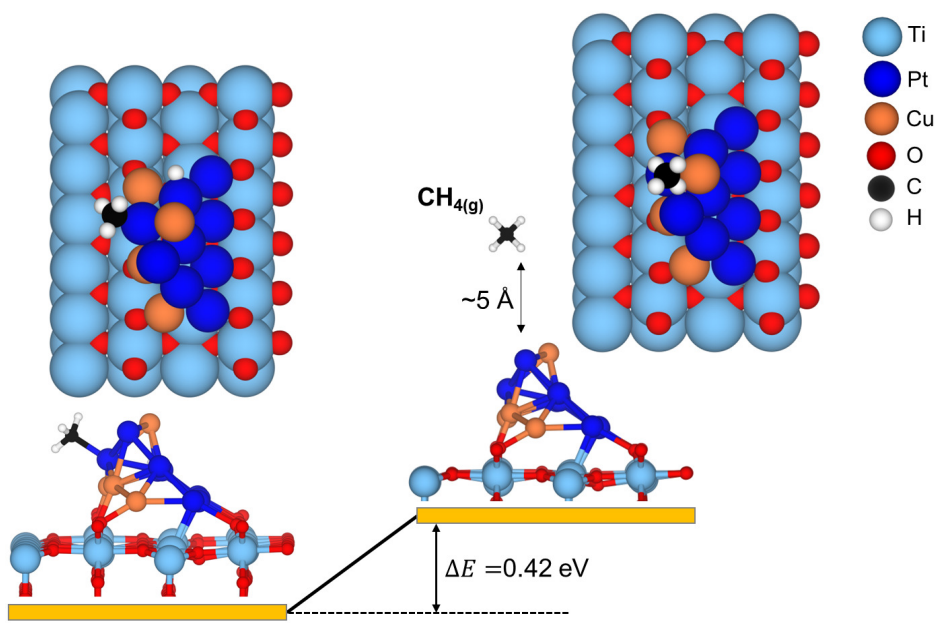

**Fig. S7**: Reaction energetics of formation of $CH_{4(g)}$ (right) from reaction between $*CH_3$ and $*H$ (left), where $CH_3$ adsorbs at a Pt site and atomic H adsorbs at another Pt site of $Cu_4Pt_9/TiO_2$. This step is endothermic with $\Delta E=0.42$ eV (more favorable than CO desorption). The top and bottom panels represent respectively the top and side views.

small perturbation to the system can induce the transfer of an atomic H from a bridging O site (bonded to Cu) to the C atom, thus resulting in the *CHOO species (Fig.S6c). For an adsorbed CHOO, C atom forms a chemical bond with a Pt atom ($d_{(C-Pt)}=2.155$ Å) and one of O atoms is bonded to both Cu and Pt atoms ($d_{(O-Cu)}=2.163$ Å and $d_{(O-Pt)}=2.178$ Å). The C-H bond length is 1.102 Å, and $O_{vac}^{ad}$-C bond length is 1.335 Å, and thus it is possible to induce the $O_{vac}^{ad}$-C bond scission. Once the $O_{vac}^{ad}$-C bond is broken, it leads to *CHO intermediate species (Fig.S6d), where CHO adsorbs at the interfacial Cu-Pt sites in a configuration, where C is bonded to a Pt site ($d_{(C-Pt)}=1.925$ Å) and O to Cu site ($d_{(O-Cu)}=2.162$ Å). The O atom of the dissociated *CHOO fills the O vacancy site.

The interfacial Cu/Pt atoms can also mediate the dissociation of $H_2O$ and can provide adsorption sites for atomic H. Indeed, H adsorbed at the interfacial Pt site (Fig.S6e) is more favorable than the *H at a bridging O atom (Fig.S6b). The Pt-H and Cu-H distances are 1.635 Å and 1.844 Å, respectively. In co-adsorption of $*CO_2$ with *H, the Pt-C distance is 2.008 Å which is slightly affected from only $*CO_2$ ($d_{(Pt-C)}=1.987$ Å). The C-$O_{vac}^{ad}$ bond length of *COO is 1.331 Å. Once it undergoes dissociation, it

5

leads to *CO at the Pt site and *O at a vacant site (Fig.S6f). The Pt-C distance is 1.841 Å, and the distance between $O^{ad}_{vac}$ (O of dissociated $CO_2$) and C atom becomes 3.411 Å. Now, the *COOH (Fig.S6g) can be formed from the reaction between $*CO_2$ and *H. We also calculated *COH structure (Fig.S6h), which might be possible between the reaction between *CO and *H. The COH adsorbs at a Pt site in a way that C atom is bonded to Pt at the interface. The Pt-C and C-O (of *COH) distances are 1.757 Å and 1.278 Å, respectively. The interaction of COH with CuPt, as mediated by the interfacial Pt atom, influences the local environment such that an interfacial Cu position, the nearest neighbor of Pt atom (COH adsorption site), is shifted in a way that it makes bonds with two bridging O atoms with Cu-O distances of 1.998 Å and 2.060 Å respectively.

### S2.6 Experiments

Fig.S8 shows the scanning electron microscopy (SEM) micrographs of two different CuPt/P25 samples, and their respective energy-dispersive X-ray spectroscopy (EDX) spectra.

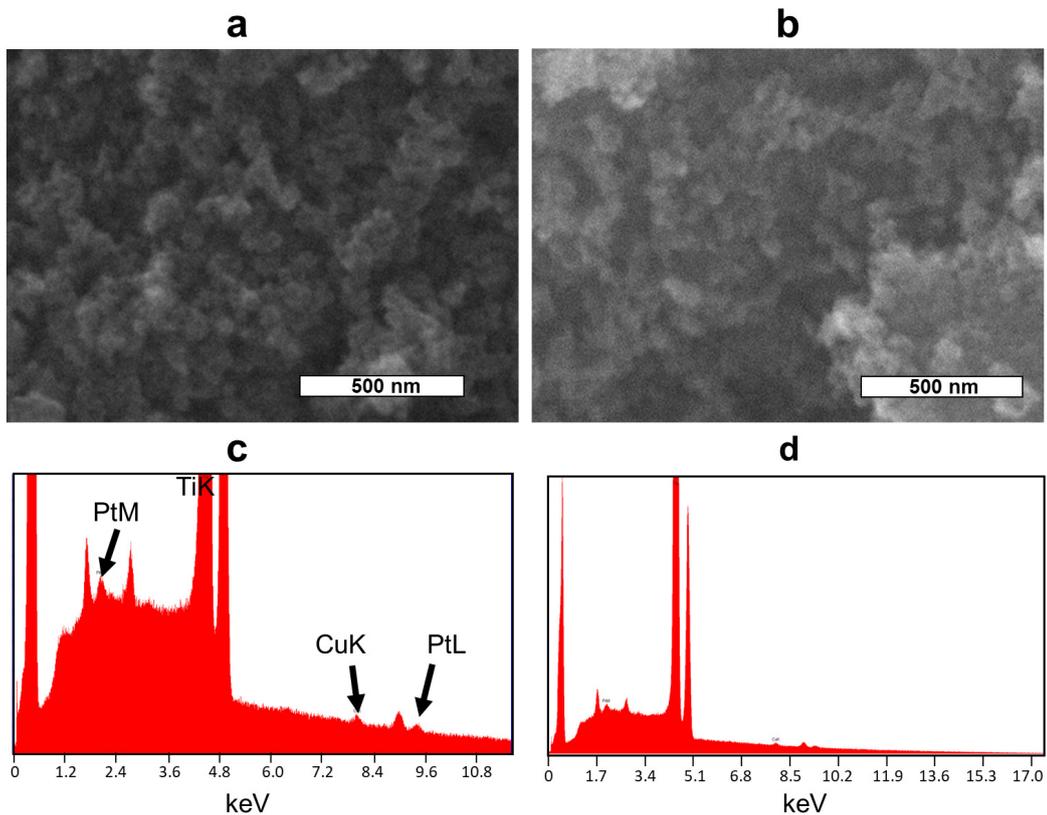

**Fig. S8**: (a-b) SEM micrographs of a) 0.14Pt-0.6Cu and b) 0.14Pt-0.4Cu samples, and (c-d) the corresponding EDX spectra respectively.

## S3 Methods

### S3.1 *Ab-initio* calculations

Some of the DFT calculations details are in the main text and some are given here. We constructed a slab system of four-layered (3×2) and (6×2)-$TiO_2$(110) systems using our calculated bulk lattice parameters (a=b=4.644 Å, and c=2.969 Å) which agree well with previous DFT studies [9, 10] and experiment [4, 11] (within ∼2%). The four and six O-Ti-O layers only differed by surface formation energy of ∼1 meV per unit area of the (110) surface, as demonstrated by the energy convergence test,[12] and therefore we chose four O-Ti-O layers. We fixed the positions of atoms of the bottom two layers of $TiO_2$(110) to ensure that those atoms behave as bulk-like (In a prior theoretical study, indeed the fixing of two bottom layers provided reliable results for $TiO_2$(110) geometry).[12] Considering several possible arrangement of Cu and Pt atoms in bimetallic Cu-Pt nanostructure, we arranged each configuration of $Cu_{(13-n)}Pt_n$

6

($n$ =0 to 13) on the (2×6) rutile $TiO_2$(110) substrate system with single O vacancy (The formation energy per oxygen vacancy was found to be 4.7 eV.[12] To avoid the spurious electrostatic interactions, we used a vacuum thickness of ∼15 Å. We relax the positions of all atoms of $Cu_{(13-n)}Pt_n$ nanoclusters and all atoms of top two O-Ti-O layers. We set the criteria of $10^{-6}$ eV for energy convergence and 0.05 eV/Å for force convergence. The Gaussian broadening of 0.1 eV was used for smearing electrons. The standard minimization algorithm was chosen for relaxations of all model structures.

### S3.2 Energetics calculations

The adsorption energy ($E_{ads}$) for adsorbates ($CO_2$, $H_2O$, and CO) is calculated as $E_{ads} = E_{adsorbate@Cu(13-n)Ptn/TiO_2} - E_{(adsorbate)} - E_{(Cu(13-n)Ptn/TiO_2)}$, where $E_{(adsorbate@Cu(13-n)Ptn/TiO_2)}$ is the DFT-calculated energy of the combined system of adsorbate and $CuPt/TiO_2$, $E_{(adsorbate)}$ is the energy of an isolated adsorbate, and $E_{(Cu(13-n)Ptn/TiO_2)}$ is the energy of $Cu_{(13-n)}Pt_n/TiO_2$.

The reaction energy (ΔE) is calculated as the difference in energy between the initial state (IS) and final state (FS) for considered each elementary step that can involve in the reaction pathways of $CO_2$ conversion process to hydrocarbons. For instance, the ΔE for elementary reaction step: *+*$CO_2$ (IS) → *CO+*O (FS), where * represents the adsorbed phase, is calculated as the difference in DFT-calculated energy of $CO_2@Cu_{(13-n)}Pt_n/TiO_2$ and that of $(CO+O)@Cu_{(13-n)}Pt_n/TiO_2$ systems.

### S3.3 Bader analysis

The charge transfer between $CO_2$ and $CuPt/TiO_2$ systems and that between CuPt and $TiO_2$ systems are estimated using the Bader decomposition of charge density based on the grid-based approach. [13, 14] The charge density is obtained self-consistently by performing the single point energy calculations of each low-energy configuration of $CO_2@Cu_{(13-n)}Pt_n/TiO_2$ systems. Using the Bader's partition scheme,[13, 14] we obtain the partial charge on each atom by the decomposition of the converged charge density. The net charge, Δq=q2-q1, (in the unit of electron ($e$)) is estimated by subtracting the reference valence charge (q1) (12$e$ for Ti; 6$e$ for O; 11$e$ for Cu; 10$e$ for Pt; and 4$e$ for C atom) from the calculated partial charge on each corresponding atom (q2).

### S3.4 Machine learning interatomic potential

We used a batch size of 5 and an initial learning rate of 0.0015, which was adaptively reduced by a factor of 0.75 upon experiencing non-decreasing loss over 10 consecutive epochs. To improve model training convergence, we employed an exponential moving average (EMA) of the model weights with decay rate 0.99 over the course of training. EMA-smoothed weights were used for the evaluation of model performance on validation and test sets. Finally, we used float32 numerical precision. While it was shown that increasing the numerical precision to float64 could improve the smoothness of the predicted potential energy surface,[15] we found that it did not significantly improve the model performance, while only incurring the high computational cost.

### S3.5 Basin-hopping Monte Carlo algorithm

The basin-hopping Monte Carlo (BHMC) algorithm uses the advantage of local minimization procedure to convert the PES from a curved surface to stepped-shaped basins.[16] The exploration of these basins was achieved by the Monte Carlo sampling through atomic displacements and the Metropolis criterion. The entropic effects on the free energy were not included while performing the basin hopping simulations using the machine learning interatomic potential to avoid adding inaccuracies. BHMC algorithm differs from the standard MC algorithm in one step namely the local optimization that is performed at each point of the PES. Since the BHMC exploration is performed by hopping among different basins, a larger atomic displacement can be used compared to standard MC. Both these features of BHMC simulations help increase the success rate in obtaining the global minima. Apart from the random displacements generally used in the MC algorithm, we also utilize custom modifiers to explore the PES thoroughly and efficiently. To explore the PES for each configuration, the following steps have been performed:

1. Locating and moving $CO_2$ adsorption site: Using a graph theory-based approach,[17] we determine various possible sites for $CO_2$ adsorption on the $TiO_2$-supported Cu-Pt nanoclusters and randomly select a site to place the adsorbate to a different location from previously chosen site.

2. Moving an oxygen vacancy position on the $TiO_2$(110) support: Bridging oxygen (2-fold coordinated) on $TiO_2$(110) plays an important role in stabilizing an adsorbed $CO_2$, as shall be discussed later. Consequently, we randomly move an O vacancy on $TiO_2$(110) to identify the stable configuration of $CO_2@Cu-Pt/TiO_2$ systems.

3. Finding mirroring plane: A randomly oriented plane is defined through the center of mass (COM) of the nanocluster. One-half of the cluster is selected and mirrored across the plane. In case the structure consists of $CO_2$, as an adsorbate on the $TiO_2$-supported Cu-Pt clusters, the mirroring is performed after removing the adsorbate followed by adding $CO_2$ on a random site found using the graph theory-based approach.[17]

4. Swapping Cu and Pt species in Cu-Pt bimetallic clusters: Randomly swapping Pt atoms with Cu and vice-versa.

The Metropolis criterion implies that a Monte Carlo move is always accepted if the free energy of a new structure, i.e. $\Delta G_{new}$, is lower than that of a previous structure, $\Delta G_{old}$, otherwise it is accepted with a probability of $\exp(\Delta G_{old} - \Delta G_{new})/k_B T_{MC}$, which is determined by a random number drawn from the interval [0,1]. Here, the Monte Carlo simulation temperature, as denoted as TMC, is an adjustable parameter. It is adjusted based on the acceptance and rejection of the structures during BHMC simulation. The flowchart of BHMC algorithm is shown in Fig.S9.

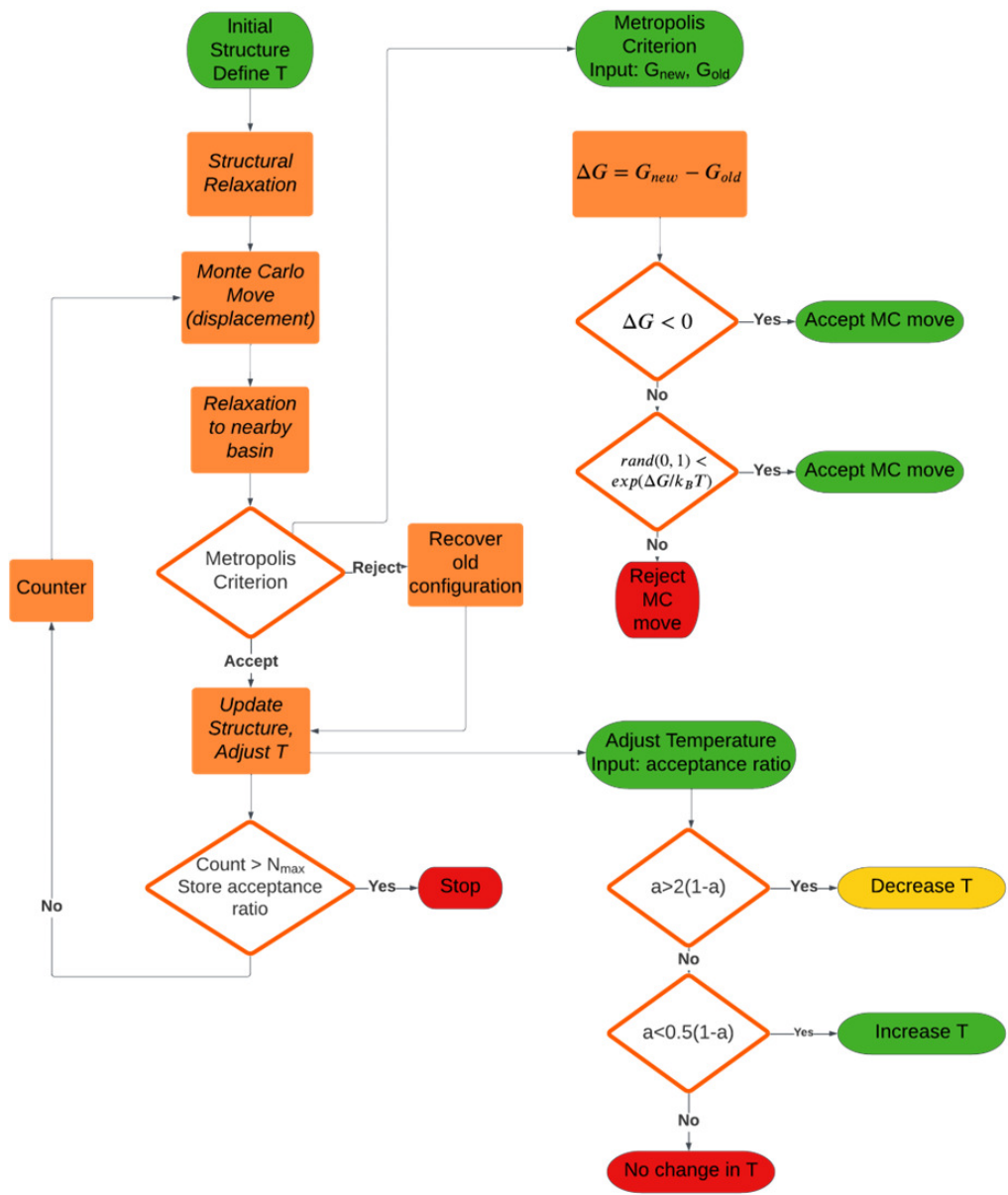

**Fig. S9**: Basin-hopping Monte Carlo algorithm flowchart.

### S3.6 Iterative training algorithm

In the iterative training process for developing the MLIP, the algorithm begins with the initial training of the potential using the available data. Subsequently, BHMC simulations are employed, utilizing the initially trained potential to explore the PES and to sample diverse atomic configurations. The diversity or uniqueness of these atomic configurations is identified by converting the generated structures data into symmetry function representations (as shall be discussed) which can be reduced in dimensionality through uniform manifold approximation and projection (UMAP)[18] and K-means clustering. Based on the clusters identified, we choose a small portion of configurations that need to be re-evaluated using DFT calculations. The new training data is generated based on the outcomes of these BHMC simulations and clustering, introducing a variety of configurations to the training set. The model is then updated and refined by incorporating the newly generated data and optimizing its parameters to better represent the complex interactions. This process iterates, with each cycle using the updated potential to perform additional simulations and continuously improving the MLIP model's accuracy. Convergence is monitored in each algorithm step, and the potential is then validated using the separate test datasets, thus ensuring generality of the algorithm applicable to several atomic configurations. Fine-tuning and optimization may be applied based on validation results, and the iteration continues until the MLIP provides satisfactory accuracy and robust generalization.

### S3.7 Symmetry function

The cutoff function takes the following form:

$$f_c(R_{ij}) = \begin{cases} 0.5 \cdot \left[\cos\left(\frac{\pi R_{ij}}{R_c}\right) + 1\right] & \text{for } R_{ij} \leq R_c \\ 0 & \text{for } R_{ij} > R_c \end{cases} \quad (1)$$

where $R_{ij}$ is the distance between atoms $i$ and $j$, and if $R_{ij}$ is larger than the cutoff radius $R_c$, the cutoff function and its derivative becomes zero.

$$G_i^2 = \sum_j e^{-\eta(R_{ij}-R_s)^2} \cdot f_c(R_{ij}) \quad (2)$$

where $\eta$ defines the width and $R_s$ defines the radial shift distance of the Gaussians.

$$G_i^4 = 2^{1-\zeta} \sum_{j,k \neq i}^{all} (1 + \lambda \cos\theta_{ijk})^\zeta \cdot e^{-\eta(R_{ij}^2 R_{ij}^2 R_{ij}^2)} \cdot f_c(R_{ij}) \cdot f_c(R_{ik}) \cdot f_c(R_{jk}) \quad (3)$$

where $\theta_{ijk} = \mathrm{acos}(\mathbf{R_{ij}R_{jk}}/R_{ik}.R_{jk})$ is the angle between atoms $j$ and $k$ with central atom $i$, $\lambda$ takes the values +1 and -1 and angular resolution is provided by $\zeta$.

The various hyperparameters for the symmetry functions are selected based on the previously used automated scheme developed by Ceriotti et al.[19]



\end{document}